\documentclass{JHEP3} 

\usepackage{graphicx}

\usepackage{amsmath}
\usepackage{amssymb}

\def\a{\alpha}

\def\e{\epsilon}

\def\L{\Lambda}
\def\r{\rho}

\def\G{\Gamma}

\def\O{\Omega}

\def\z{\zeta}

\def\ra{\rightarrow}
\def\Ra{\Rightarrow}

\def\Mfunction#1{\mathop{\rm #1}\nolimits}

\title{Fractal analysis of the dark matter and gas distributions in the
Mare-Nostrum universe} 
\author{Jos\'e Gaite\\ Instituto de Microgravedad IDR,
EIAE,\\ Universidad Polit\'ecnica de Madrid, E-28040 Madrid, Spain;\\
jose.gaite@upm.es }

\preprint{\today}

\abstract{ We develop a method of multifractal analysis of $N$-body
cosmological simulations that improves on the customary counts-in-cells method
by taking special care of the effects of discreteness and large scale
homogeneity.  The analysis of the Mare-Nostrum simulation with our method
provides strong evidence of self-similar multifractal distributions of dark
matter and gas, with a halo mass function that is of Press-Schechter type but
has a power-law exponent $-2$, as corresponds to a multifractal.  Furthermore,
our analysis shows that the dark matter and gas distributions are
indistinguishable as multifractals.  To determine if there is any gas biasing,
we calculate the cross-correlation coefficient, with negative but inconclusive
results. Hence, we develop an effective Bayesian analysis connected with
information theory, which clearly demonstrates that the gas is biased in a
long range of scales, up to the scale of homogeneity.  However, entropic
measures related to the Bayesian analysis show that this gas bias is small (in
a precise sense) and is such that the fractal singularities of both
distributions coincide and are identical. We conclude that this common
multifractal cosmic web structure is determined by the dynamics and is
independent of the initial conditions.}


\begin{document}

\section{Introduction}

The large scale structure of the Universe can be described as a ``cosmic web''
formed by matter sheets, filaments and nodes.  This type of structure was
initially proposed in connection with simplified but insightful models of the
cosmic dynamics \cite{Shan-Zel} and has been since confirmed by galaxy surveys
and $N$-body cosmological simulations \cite{Rien}.  Cosmological simulations
have been especially helpful in testing models of structure formation. In a
sense, they have been complementary to observations, since observations are
biased towards the luminous matter, while simulations have fully considered
the evolution of the dark matter, which is actually the dominant component. In
fact, many simulations {\em only} consider dark matter, in particular,
non-baryonic cold dark matter, whose dynamics is simplest to simulate and
gives rise to cosmic structure that is in accord with observations.  However,
due to the advances in parallel computing, the development of efficient codes,
and the availability of more powerful computers, the scope of $N$-body
simulations has recently changed: now it is possible to simulate the combined
dynamics of the non-baryonic dark matter and the baryon gas in large
cosmological volumes and with relatively good resolution.

We analyse here the data output of a recent large cosmological simulation of
the combined dark matter and gas dynamics, namely, a simulation of the cosmic
evolution of $1024^3$ dark-matter particles and an equal number of gas
particles carried out by the Mare-Nostrum supercomputer in Barcelona. This
dataset has already been analysed by the researchers in charge of the
Mare-Nostrum universe project \cite{Gott1,Falten,Gott2}.  Here, we are
interested in a particular aspect of the dark matter and gas distributions:
their geometry and, specifically, their fractal geometry.

Fractal geometry \cite{Mandel} is the geometry of sets or distributions that
have noticeable geometrical features on ever decreasing scales. It is related
to scale invariance and indeed appears in nonlinear dynamical systems in which
the dynamics is characterized by the absence of reference scales. This is the
case of the dynamics of collision-less cold dark matter (CDM), only subjected
to the gravitational interaction.  Therefore, the cosmic web produced by this
type of dynamics has fine structure and it is, arguably, statistically
self-similar. We can reasonably assume that the cosmic web is a multifractal
attractor of the gravitational dynamics.  This model is supported by the
results of CDM simulations \cite{Valda,Colom,Yepes,I4,I5}.  Although the gas
dynamics is more complex (due to the gas pressure, etc), the gas takes part in
the nonlinear dynamics of structure formation and can also have a multifractal
attractor. Indeed, scaling laws in the distribution of galaxies have a long
history, which has been reviewed in
Refs.~\cite{Borgani,Jones-RMP,Sylos-Pietro}.  Therefore, it is interesting to
compare the scaling laws in the distribution of gas with the scaling laws in
the distribution of dark matter.

Fractal models of the cosmic structure can only be valid in a range of scales,
whose upper cutoff is the scale of homogeneity. Its value has been the subject
of considerable debates and still is controversial
\cite{Jones-RMP,Sylos-Pietro}. In contrast, the lower cutoff to scaling has
attracted less attention.  In fact, the CDM gravitational dynamics does not
introduce any small reference scale that can play the r\^ole of a lower
cutoff, but the gas dynamics introduces the Jeans length.  This length is not
a fixed reference scale, for it depends on the local thermodynamical
parameters.  In any event, one should expect that the lower cutoff to scaling
in the dark matter distribution is smaller than the lower cutoff appropriate
for the distribution of galaxies.  However, the opposite seems to be true if
one compares galaxy surveys with the results of cosmological simulations,
since the latter exhibit reduced scaling ranges, even in dark matter only
simulations.  Peebles has included this problem in his list of anomalies in
standard cosmology \cite{Pee}.  In his words: ``scale-dependent biasing seems
an awkward way to account for the power-law forms of the low order galaxy
position correlation functions.''

One can be inclined to place more trust in the scaling range found in galaxy
surveys: cosmological simulations allow one to obtain better statistics but
they are not free of systematic errors that affect an important range of the
smaller scales. Indeed, it has been long known that $N$-body simulations are
not fully reliable on scales smaller than the mean particle spacing $N^{-1/3}$
\cite{KMS,KMSS}.  In spite of the ever-growing value of $N$, the range of
scales between the scale $N^{-1/3}$ and the homogeneity scale is still rather
small.  In the Mare-Nostrum universe, this scale range spans a factor of 30
(see Sects.~\ref{anal} and \ref{MF}).  Our goal is to demonstrate
multifractality of the dark matter and gas distributions in the valid scale
range.  Furthermore, given that this scale range is small, we devise a method
to correct for discreteness effects and thus extend the valid range to smaller
scales, obtaining a reasonable scaling range.  We also intend to test if the
dark matter and gas distributions constitute a unique distribution or to what
extent they differ. Hence, we make a model of {\em fractal biasing}.

We describe our method of coarse multifractal analysis by counts in cells and
define the basic objects (halos) in Sect.~\ref{anal}.  In our method, the
scale of homogeneity is explicitly introduced to calculate the multifractal
spectrum (Sect.~\ref{anal_w_homo}).  In Sub-sect.~\ref{features}, we show how
to obtain the main features of this spectrum and how they are influenced by
discreteness and large scale homogeneity.  In Sect.~\ref{MF}, we apply our
method to the zero-redshift particle distributions of the Mare-Nostrum
universe: (i) we obtain the halo mass functions and discuss its relation to
the Press-Schechter mass function in Sect.~\ref{mfun}; (ii) we obtain the
multifractal spectra and discuss their relevance in regard to other
geometrical studies of the cosmic web in Sect.~\ref{MFsp}; and (iii) we
demonstrate scaling and compute sound values of the correlation dimensions in
Sect.~\ref{frac_dist}.  The similarity of the results corresponding to the gas
and the dark matter suggests that both distributions are identical and shows
the need of precise statistical methods to discriminate between them
(Sect.~\ref{bias}).  Since the cross-correlations cannot give a definite
answer (Sub-sect.~\ref{cross}), we develop an effective Bayesian analysis
(Sub-sect.~\ref{Bayes-sect}) which we apply to various cell distributions
(Sub-sect.~\ref{appl_Bayes}). This analysis connects with the thermodynamic
entropy of mixing (Sub-sect.~\ref{G-DM_entropy}).  Therefore, we study the
application of entropic measures to discriminating between mass distributions,
and we study the connection of entropies in the continuum limit with the
multifractal spectrum (Sect.~\ref{entropies}).  Finally, we discuss our
results (Sect.~\ref{discuss}).

A note on notation: we use frequently the asymptotic signs $\sim$ and
$\approx$; for example, $f(x) \sim g(x)$ or $f(x) \approx g(x)$ (often without
making explicit the independent variable $x$). The former means that the limit
of $f(x)/g(x)$ is finite and non-vanishing when $x$ approaches some value
(which can be zero or infinity), while the latter means, in addition, that the
limit is one.  We also use the sign $\simeq$, which only refers to imprecise
numerical values (with unspecified errors).

\section{Methods of data analysis}
\label{anal}

The Mare-Nostrum cosmological simulation is described by Gottl\"ober et al
\cite{Gott1}.  It assumes a spatially flat concordance model with parameters
$\O_\L=0.7$, $\O_{\bar{\rm m}}=0.3$, $\O_{\rm bar}=0.045$, Hubble parameter $h
= 0.7$, and initial spectrum with spectral index $n=1$, in a comoving cube of
500 $h^{-1}$ Mpc edges.  The Gadget-2 code \cite{Springel} simulated the
evolution of dark matter and gas from redshift $z=40$ to $z=0$. Both dark
matter and gas are resolved by $1024^3$ particles, respectively, which results
in a mass of $8.24\cdot 10^9\,h^{-1}\, M_\odot$ per dark-matter particle and a
mass of $1.45\cdot 10^9\,h^{-1}\, M_\odot$ per gas particle. The Gadget-2 code
implements polytropic (adiabatic) evolution of the gas. It can also include
dissipation due to radiation or conduction, but these processes have not been
included in the Mare-Nostrum simulation.  Nevertheless, the code always
includes an artificial viscosity to take care of shock waves.

The Mare-Nostrum universe consists of 135 evenly spaced snapshots.  For our
statistical analysis, we only need the $z=0$ snapshot, in which the
homogeneity scale is largest and the structures are most developed.  The large
size of a Mare-Nostrum universe snapshot makes it unwieldy, so it is
convenient (and almost necessary) to analyse it in terms of compound
structures, namely, halos, rather than analysing the full particle
distributions.  The Mare-Nostrum universe researchers
\cite{Gott1,Falten,Gott2} use a friends-of-friends algorithm to define halos,
and then they study the distribution and features of those halos. However, we
prefer the method of counts in cells, more suitable for studying the continuum
limit and the scaling properties of particle distributions.  Therefore, our
elementary objects (halos) are cells with constant size but variable mass. The
definition of elementary objects in distributions with fine structure
(fractals) is arbitrary to a high degree, being actually tied to the
measurement or analysis technique.  The definition of elementary objects by
coarse graining and, in particular, their definition as cells in a mesh, is
very convenient \cite{I4}.  In absence of a reference scale, the appropriate
cell size (the coarse-graining scale) is arbitrary, and there is no clear
distinction between inner and outer structure.  However, an $N$-point fractal
sample, as a {\em finite} point distribution, has a reference scale, namely,
the discreteness scale $N^{-1/3}$, which allows us to properly define the
size of elementary objects.

At any rate, the cell size must be considered a running scale.  The use of a
running cell size is useful, for example, to distinguish the nonlinear scales
where structure formation takes place from the linear scales where the initial
conditions are preserved: as the cell size enters in the range of the latter
scales, the fluctuations of the counts in cells are reduced to small Gaussian
fluctuations. This homogeneity scale is actually the only real scale in the
cosmic CDM dynamics, although it is not a sharp scale and, besides, it grows
with time.

The method of counts in cells is also suitable for comparing the gas
distribution with the dark matter distribution, by comparing the respective
counts, for a given cell size.  Of course, we must devise methods to provide
these comparisons with statistical meaning.  We defer further description of
our methods to Sect.~\ref{bias}. However, we advance that our main procedure
naturally connects with the description of multifractals in terms of R\'enyi
dimensions.

In summary, our basic assumption is that the Mare-Nostrum particle
distributions represent continuous mass distributions with fine structure but
which are homogeneous on the large scales.  In particular, we expect
continuous distributions of cosmic web type, which have various kinds of
density singularities produced by gravitational collapse.  The properties of
these singularities can be deduced by suppressing the effects of discreteness.
We introduce in next sub-section methods of multifractal analysis
geared to the relevant type of singular distributions. In sub-section
\ref{features}, we study the influence of the discreteness scale and the
homogeneity scale on the features of the coarse multifractal spectrum.

\subsection{Counts in cells and coarse multifractal analysis}
\label{anal_w_homo}

Let us assume that a mesh of cells is placed in the sample region (the
simulation cube).  In the method of counts in cells, (fractional) statistical
moments are defined as
\begin{equation}
M_q = \sum_i \left(\frac{n_i}{N}\right)^{q} = 
\sum_{n>0} N(n)\left(\frac{n}{N}\right)^{q},
\label{Mq}
\end{equation}
where the index $i$ refers to non-empty cells, $n_i$ is the number of points
(particles) in the cell $i$, $N= \sum_i n_i$ is the total number of points,
and $N(n)$ is the number of cells with $n$ points.%
\footnote{Central moments are defined by subtracting from $n/N$ its
average. In the strongly nonlinear regime, central moments are less
convenient.}  The second expression involves a sum over cell populations and
it is more useful than the sum over individual cells, because the range of $n$
is much smaller (when the cell size is small).  $M_0$ is the number of
non-empty cells and $M_1 = 1$. We understand the latter as a mass
normalization, namely, the mass in cell $i$ is $n_i/N$ and the total mass is
one, such that the mass distribution can be interpreted as a probability
distribution (the physical masses of gas or dark-matter particles play no
r\^ole in the statistical analysis).  There is an alternate definition of
$q$-moments:
\begin{equation}
\mu_q = \langle \r^q \rangle 
= \sum_{n>0} \frac{N(n)}{M_0} \left(\frac{n}{NV}\right)^{q}
= \frac{M_q}{V^q M_0} \,,
\label{muq}
\end{equation}
where $V$ is the cell's volume, $N(n)/M_0$ is the fraction of cells that
contain $n$ points, and $\r=n/(NV)$ is the density in those cells. With this
definition, $\mu_0 =1$ while $\mu_1$ is not fixed.  We notice that the moments
with positive integer $q$ ($M_q$ or $\mu_q$, $q \in \mathbb{N}$) are
sufficient for regular distributions, but we cannot impose this restriction
here ($q \in \mathbb{R}$).

In regular distributions, the mass contained in any cell is proportional to
its volume $V$, in the continuum limit $V \ra 0$. Therefore, $M_q \sim
V^{q-1}$.  However, we consider singular distributions such that their
$q$-moments are non-trivial power laws of $V$ in the continuum limit, namely,
distributions such that one can define \cite{Harte} the exponents
\begin{equation}
\tau(q) = 3\lim_{V\ra 0}\frac{\log M_q}{\log V}\,,\;q \in \mathbb{R}\,.
\label{tauq}
\end{equation}
These distributions are called multifractals.%
\footnote{The mathematical definition of a multifractal distribution only
  requires the existence of $\tau(q)$, which is a mild condition on the type
  of singularities and does not necessarily imply self-similarity. For
  example, an isolated power-law singularity or a massive particle in a
  uniform background both give rise to non-trivial ``bifractal'' functions
  $\tau(q)$.  Nonetheless, physically relevant distributions with non-trivial
  $\tau(q)$ usually exhibit some kind of self-similarity, albeit in a
  statistical sense.}
Of course, the numerical evaluation of the limit in Eq.~(\ref{tauq}) is not
feasible and one must be satisfied with finding a constant value of the
quotient for sufficiently small $V$, that is, in a sufficiently long range of
negative values of $\log V$ (a range of scales).  In fact, the exponent is
normally defined as the slope of the plot of $\log M_q$ versus $\log V$, and
its value is found by numerically fitting that slope, supposing that a
meaningful fit is possible.

A multifractal is also characterized by a set of {\em local} dimensions: the
local dimension at one point says how the mass grows from that point
outwards. Every set of points with a given local dimension $\a$ constitutes a
fractal set with dimension $f(\a)$.  In terms of $\tau(q)$, the spectrum of
local dimensions is given by
\begin{equation}
\a(q)= \tau'(q)\,,\quad q \in \mathbb{R}\,,
\label{aq}
\end{equation}
and the spectrum of fractal dimensions $f(\a)$ is given by the Legendre
transform
\begin{equation}
f(\a) = q\,\a - \tau(q)\,.
\label{fa}
\end{equation}
The spectrum of fractal dimensions is convex upwards and fulfills $f(\a) \leq
\a$.  The fractal dimension $f(\a)$ reaches the local dimension $\a$ at $q=1$
[note that Eq.~(\ref{tauq}) gives $\tau(1)=0$].  The set of singularities with
$f(\a_1) = \a_1$ contains the bulk of the mass and is called the ``mass
concentrate.''

In addition to the exact exponent $\tau(q)$ (\ref{tauq}), we define, for a
given cell size, the {\em coarse} exponent
\begin{equation}
\tau(q) = 3\frac{\log (M_q/V_0^{q-1})}{\log (V/V_0)}\,,
\label{ctauq}
\end{equation}
where $V$ is the cell size and $V_0$ is the homogeneity scale, such that the
density is homogeneous and $M_q \approx V^{q-1}$ for $V > V_0$.  The coarse
exponent depends on both $V$ and $V_0$, but this dependence vanishes if $V \ll
V_0$ (assuming that the limit $V \ra 0$ exists).  The introduction of the
homogeneity scale in Eq.~(\ref{ctauq}) improves the definition used in
Ref.~\cite{I4} for the GIF2 simulation, where no $V_0$ is introduced
(equivalent to setting $V_0 = 1$).  Given that the Mare-Nostrum universe cube
has 500 $h^{-1}\!$ Mpc edges, much longer than the 110 $h^{-1}\!$ Mpc edges of
the GIF2 simulation cube, it is important now to take the transition to
homogeneity into account in the definition of the coarse exponent, if we want
it to be a good approximation of the limit (\ref{tauq}) for moderately small
$V$.

The homogeneity scale $V_0$ can be found as the scale of crossover to
homogeneity in the scaling of statistical moments (Sect.~\ref{frac_dist}).  We
can also estimate it as the coarse-graining scale such that the mass
fluctuations are smaller than, say, 10\%; namely, we define it as the scale
such that $\mu_2 = 1.1$.  Thus, we find that the scale of homogeneity is about
1/16th of the edge of the cube, namely, about 30 $h^{-1}$ Mpc.  This value is
similar to the value of the GIF2 homogeneity scale found in Ref.~\cite{I4},%
\footnote{The value found in Ref.~\cite{I4},
$r_0 \simeq 14\, h^{-1}\!$ Mpc, is
roughly equivalent to {\em half} the edge of the cube such that $\mu_2 <
1.1$.}  
where it is calculated from the crossover in the scaling of moments.

Besides the multifractal spectrum $f(\a)$, it is useful to define the spectrum
of R\'enyi dimensions \cite{Harte}
\begin{equation}
D_q= \frac{\tau(q)}{q-1}\,.
\label{Dq}
\end{equation}
They have an information-theoretic meaning, which will be explained in detail
in Sect.~\ref{entropies}.  In particular, the dimension of the mass
concentrate $\a_1 = f(\a_1) = D_1$ is also called the entropy dimension.
$D_0$ coincides with the maximum value of $f(\a)$ and with the box-counting
dimension of the distribution's support, while $D_2 = \tau(2)$ is the
correlation dimension.  In the homogeneous regime, $M_q \approx V^{q-1}$ and
$D_q= 3$ for any $q$.  In a uniform fractal (a {\em unifractal} or {\em
monofractal}) $D_q$ is also constant but smaller than three. In general, $D_q$
is a non-increasing function of $q$.

\subsection{Features of the coarse multifractal spectrum}
\label{features}

Here, we examine the features of the multifractal spectrum obtained from the
coarse exponent defined by Eq.~(\ref{ctauq}).  

In a multifractal, the cell size $V$ is, of course, irrelevant, as long as $V$
is sufficiently smaller than the homogeneity scale $V_0$. However, the
intrinsic discreteness of a multifractal sample (a finite point distribution)
gives rise to another scale, namely, the size of the cell such that there is
one point per cell on average ($V =N^{-1}$).  This scale represents the
minimal scale at which the distribution can be consistently considered
continuous.  In the initial stages of an $N$-body simulation, when there are
only very small deviations from the one-particle-per-cell average, it is
obvious that it makes no sense to consider smaller scales.  Furthermore, the
dynamics of gravitational collapse is deeply distorted on volumes $V <
N^{-1}$, so the resulting particle clusters do not represent the structures
that result from the collapse of a continuous medium \cite{KMS,KMSS}.
As a coarse-graining scale, the volume $V =N^{-1}$ produces
the largest variety of masses of coarse-grained objects in $N$-body
cosmological simulations \cite{I4}.  Thus, this cell size provides us with a
sort of {\em master cell distribution} that characterizes the multifractal
sample.  Whenever we mention halos, we refer to non-empty master cells,
preferably with a considerable number of particles.  Since the number of
dark-matter or gas particles in the Mare-Nostrum universe is a perfect cube
and, indeed, a power of two, the master cell distributions are easily
obtained.

Ref.~\cite{I4} shows that the mass function of halos in the GIF2 simulation
follows the power law $N(m) \sim m^{-2}$, except at the large mass end, where
it decays faster.  This power law derives from an approximation of the
multifractal spectrum, namely, $f(\a) \approx \a$, and therefore represents
the mass concentrate of the multifractal.  In contrast, the master cell
distribution contains no information of the matter distribution in voids
(zones with $\a > 3$), because they are empty \cite{I4,I5}. Hence, a part of
the multifractal spectrum is missing even at this scale.  As $V$ shrinks, the
multifractal spectrum is reduced further.

The length scale that corresponds to $V =N^{-1}$ in the Mare-Nostrum
simulation, namely, $l = N^{-1/3} = 2^{-10}$, is only a factor $2^{6}=64$
smaller than $l_0 = V_0^{1/3} = 2^{-4}$.  This is the largest scaling range
that could be attainable in principle, despite the large number of particles.
In fact, close to the large scale end, at $l_0 = 2^{-4}$, the coarse
multifractal spectrum is influenced by homogeneity, whereas close to the
opposite end it is influenced by discreteness. Surely, the best estimation of
the real spectrum is to be found somewhere in between. Let us study in detail
the change of the features of the coarse multifractal spectrum with scale.

For a given coarse-graining scale, we calculate with Eqs.~(\ref{Mq}) and
(\ref{ctauq}) the exponent $\tau(q)$, and hence we calculate the coarse
multifractal spectrum through the Legendre transform given by (\ref{aq}) and
(\ref{fa}).  The lower end of this spectrum corresponds to the limit $q\ra
\infty$, that is to say, to the cell(s) with maximum number of particles:
\begin{eqnarray}
\a_{{\rm min}}=\lim_{q\ra \infty}\a(q) = 3\frac{\log [n_{{\rm
max}}/(NV_0)]}{\log (V/V_0)}\,,
\label{amin}
\\ 
f(\a_{{\rm min}}) = -3\frac{\log [N(n_{{\rm max}})\,V_0]}{\log (V/V_0)}\,.
\label{famin}
\end{eqnarray}
Since $\a_{{\rm min}}$ is the local dimension of the strongest singularity, it
changes little with the scale, unless we approach homogeneity ($V \to V_0$),
which implies that $\a_{{\rm min}} \to 3$.  Usually, $N(n_{{\rm max}})=1$,
namely, there is only one cell with the maximum number of
particles. Therefore, the choice $V_0=1$, which disregards the effect of
homogeneity, implies that $f(\a_{{\rm min}}) =0$. However, any $V_0 < 1$, like
our present setting $V_0 = 1/4096$, implies that the fractal dimension
$f(\a_{{\rm min}})$ is negative!

Intuitively, negative fractal dimensions seem meaningless, but they often
arise in the study of random multifractals. The origin of negative fractal
dimensions has been discussed by Mandelbrot \cite{Mandel2}.  In brief, the
coarse fractal dimension of a set of singularities in a random multifractal is
proportional to the logarithm of their number, but the expected value of this
number can be smaller than one. Therefore, sets of singularities with negative
fractal dimension are probably empty.  In our case, by setting $V_0$ to a
fraction of the total volume, the number of singularities with given $\a$ in
cubes of size $V_0$ fluctuates and these fluctuations are more important for
values of $\a$ such that there are few singularities with that $\a$ in the
whole simulation box.  Thus, it is convenient to ``average'' over the
$V_0^{-1}=4096$ cubes and consider at once the 4096 singularities with
smallest $\a$, truncating the negative values of the multifractal spectrum.

In analogy with the lower end of the spectrum of local dimensions, we can
deduce that its upper end corresponds to the limit $q\ra -\infty$, that is, to
the set of cells with one particle (assuming that $V$ is not so large that
there are none). In fact,
\begin{eqnarray}
\a_{{\rm max}}=\lim_{q\ra -\infty}\a(q) = -3\frac{\log (N V_0)}{\log(V/V_0)}\,,
\label{amax}
\\
f(\a_{{\rm max}}) = -3\frac{\log [N(1) V_0]}{\log (V/V_0)}\,.
\label{famax}
\end{eqnarray}
Notice that the master cell distribution has $\a_{{\rm max}}=3$ and,
therefore, its spectrum is limited to non-void zones ($\a \leq 3$).  The value
of $\a_{{\rm max}}$ increases for cell sizes $V > 1/N$, as voids begin to be
sampled.  For sufficiently large $V$, $N(1)$ decreases and approaches $1/V_0 =
4096$ (only one cell with one particle per each cube of size $V_0$).  Then,
$f(\a_{{\rm max}})$ decreases to zero.  At this scale, we have the complete
(positive) multifractal spectrum in the region $\a > 3$, corresponding to
voids, and the distribution can be considered continuous over the entire range
of $\a$ [we always discard the negative values of $f(\a)$].

The total span of the spectrum is
$$ \a_{{\rm max}} - \a_{{\rm min}} = -3\frac{\log (n_{{\rm max}}/n_{{\rm
min}})}{\log (V/V_0)}\,,
$$ 
where $n_{{\rm min}} \equiv 1$ in the relevant range of $V$.  Naturally, the
largest span is reached when the spectrum is complete in the region $\a > 3$.
For the Mare-Nostrum universe, we indeed show in Sect.~\ref{MFsp} that we
obtain, by choosing $V$ to be the largest value such that $f(\a_{{\rm max}})
\geq 0$, the largest span of dimensions $\a$ and a good estimate of the full
multifractal spectrum.  For larger values of $V$, as the transition to
homogeneity begins, $n_{{\rm min}}$ grows and approaches $n_{{\rm max}}$, with
the consequent contraction of the span of the spectrum.

Scale invariance implies that the multifractal spectra at different
coarse-graining scales coincide in their respective ranges $[\a_{{\rm
min}},\a_{{\rm max}}]$, where $\a_{{\rm min}}$ is roughly constant but
$\a_{{\rm max}}$ increases with the scale.  However, the under-sampling of low
density regions that causes the truncation of the spectrum at $\a_{{\rm max}}$
also causes deviations from the true spectrum close to $\a_{{\rm max}}$. These
deviations must be corrected. We see how to do it for the Mare-Nostrum
multifractal spectra in Sects.~\ref{MFsp} and \ref{frac_dist}.

Regarding the master cell distribution and assuming for it the simple mass
function $N(n) = N(1)/n^2$, we can deduce interesting consequences about the
corresponding coarse multifractal spectrum.  First, we calculate, according to
Eq.~(\ref{Mq}),
$$
M_0 = \sum_{n=1}^{n_{{\rm max}}} N(n) \approx N(1) 
\sum_{n=1}^{\infty} \frac{1}{n^2} = N(1)\,\frac{\pi^2}{6}\;.
$$ 
Since this sum is just the number of non-empty cells, we deduce that the
fraction of non-empty cells containing one particle is $N(1)/M_0 \approx
6/\pi^2 = 0.61$.  Thus, the full distribution $N(n)$ is determined by just the
number of empty cells.  Furthermore, from the expression
\begin{equation}
M_1 = \sum_{n=1}^{n_{{\rm max}}} N(n) \frac{n}{N} \approx 
\frac{N(1)}{N} \ln n_{{\rm max}}\;,
\label{M1}
\end{equation}
and the condition $M_1 \equiv 1$ we can determine $n_{{\rm max}}$.  Then, the
dimension of the mass concentrate $\a_1 = f(\a_1)$ is
\begin{eqnarray*}
\a_1 &=& \tau'(1) = \frac{3}{\ln (V/V_0)} 
\left(\left.\frac{dM_q}{dq}\right|_{q=1} - \ln V_0 \right) \\
&=& \frac{3}{\ln (V/V_0)}\left( \sum_{n=1}^{n_{{\rm max}}} N(n) \frac{n}{N} 
\ln \frac{n}{N} - \ln V_0 \right) \\
&\approx& \frac{3}{\ln (V/V_0)}\,
\left(\frac{\ln n_{{\rm max}}}{2} - \ln (NV_0) \right). 
\end{eqnarray*}
This dimension is the arithmetic mean of the general values of $\a_{{\rm
min}}$ in Eq.~(\ref{amin}) and $\a_{{\rm max}}$ in Eq.~(\ref{amax}).

\section{Multifractal analysis of the dark matter and gas distributions}
\label{MF}

We now present the results of the multifractal analysis of the Mare-Nostrum
universe $z=0$ snapshot, beginning with the halo mass functions given by the
counts in the master cell distributions (Sect.~\ref{mfun}).  In
Sect.~\ref{MFsp}, we study the multifractal spectra in the range of scales
covering several powers of two, namely, from $l = 2^{-12}$ to $l =
2^{-7}$. The latter scale is the smallest scale (among the powers of two) such
that $N(1) < 4096$ and therefore the spectrum corresponding to voids is
complete.  On smaller scales, namely, between $l = 2^{-12}$ and $l = 2^{-8}$,
the high-$\a$ ends of the coarse spectra deviate from the true spectrum due to
under-sampling of the low density regions.  In Sect.~\ref{frac_dist}, we
propose to correct for under-sampling by removing the erroneous ends of the
spectrum.  Thus, we can demonstrate scale invariance in the longest possible
range.

\subsection{Mass functions}
\label{mfun}

\begin{figure}
\centering{\includegraphics[width=7.5cm]{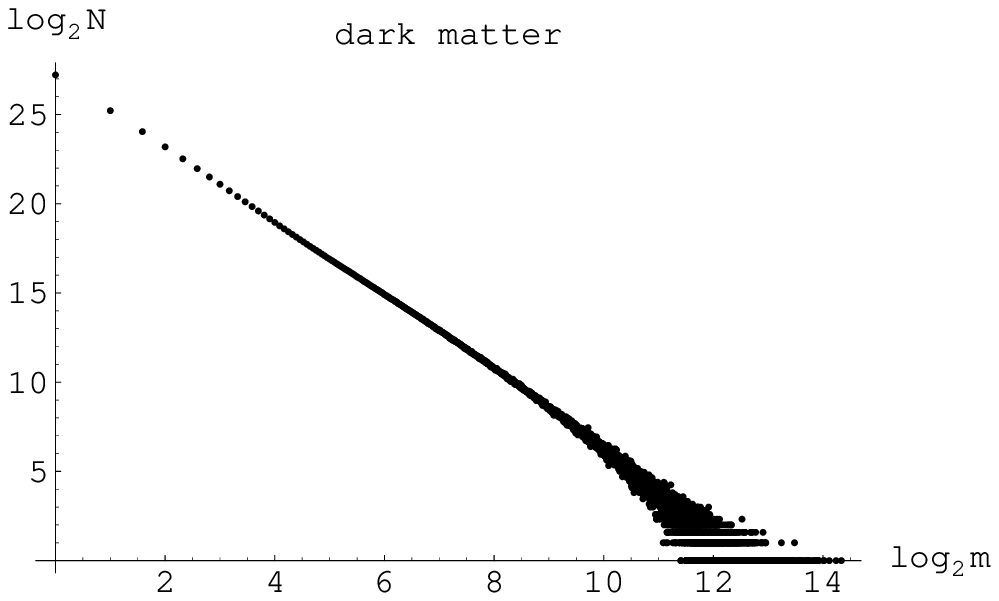}}
\centering{\includegraphics[width=7.5cm]{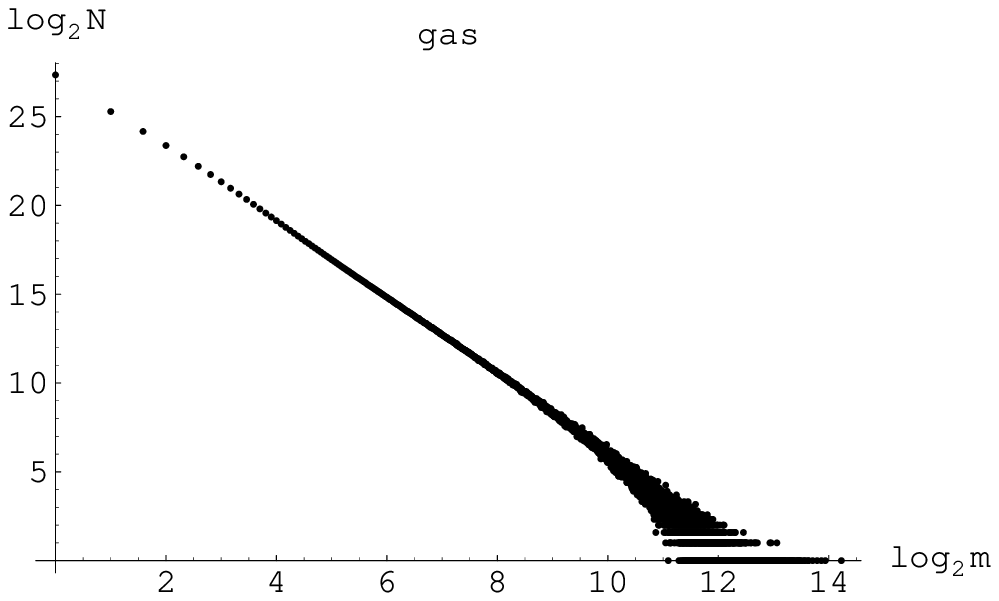}}
\caption{Log-log plots of the number of halos $N$ versus
their mass $m$ (number of particles)
at coarse-graining scale $1024^{-1}$, in the case of dark matter 
(left) and gas (right).
}
\label{P-S}
\end{figure}

In Fig.\ \ref{P-S} are plotted the halo mass functions of dark-matter and gas,
obtained from the counts in the master cell distributions.  The mass $m$ is
actually defined as the number of particles, for simplicity.  Both mass
functions follow the power law $N(m) \sim m^{-2}$ over a considerable range of
$m$: least-squares fits in the $\log_2 m$ range from 0 to 9 yield slopes
$-2.07$, for the dark matter, and $-2.12$, for the gas.

There are $156\hspace{1pt}272\hspace{1pt}463$ cells with one dark-matter
particle and $170\hspace{1pt}546\hspace{1pt}782$ cells with one gas particle
in the master cell distribution.  According to Eq.~(\ref{famax}), the fractal
dimensions of the sets with $\a_{{\rm max}}=3$ are $f(\a_{{\rm max}})=2.54$
and $2.56$, for the dark matter and gas, respectively.  The cell with the
largest proportion of dark matter has $20\hspace{1pt}658$ dark-matter
particles and it also has the largest proportion of gas, namely,
$19\hspace{1pt}200$ gas particles; all the particles together form the most
massive halo.  The corresponding values of $\a_{{\rm min}}$, according to
Eq.~(\ref{amin}), are $0.61$ and $0.63$, respectively. However,
Eq.~(\ref{famin}) yields negative values of $f(\a_{{\rm min}})$, which we do
not consider.  We compute directly from Eqs.~(\ref{aq}), (\ref{fa}) and
(\ref{ctauq}) that the values of $\a$ such that $f(\a) = 0$ are $0.91$ (dark
matter) and $0.95$ (gas).

We have seen in the preceding section that the value of $\a_1$ corresponding
to the master cell distribution can be estimated as the arithmetic mean of
$\a_{{\rm min}}$ and $\a_{{\rm max}}=3$.  Whether we use $\a_{{\rm min}}
\simeq 0.6$ or $\a_{{\rm min}} \simeq 0.9$, this estimation yields smaller
values than the actual values, which are $2.22$ (dark matter) and $2.29$
(gas).  On the other hand, the estimation $m_{{\rm max}} = \exp[N/N(1)]$,
deduced by making $M_1 =1$ in Eq.~(\ref{M1}), yields 967 and 542,
respectively, well below the real values (see Fig.\ \ref{P-S}).  The problem
is that the power law is modified at the large mass end, as we can perceive in
Fig.\ \ref{P-S}. On the one hand, at the large mass end, the values of $N(m)$
are so small that there are many values of $m$ for each value of $N$; on the
other hand, $N$ as a function of the average of the corresponding values of
$m$ decays faster than a power law.  In fact, the above estimated values of
$m_{{\rm max}}$ actually mark the ends of the power laws, instead of the ends
of the large masses.

We can improve the fit of the mass function by modelling the large mass end of
the power law. For this, we can take inspiration from the Press-Schechter mass
function,
\begin{equation}
N(m) \propto \left(\frac{m}{m_*}\right)^{n/6 - 3/2}
\exp\left[-\left(\frac{m}{m_*}\right)^{n/3 + 1}\right],
\label{P-S_MF}
\end{equation}
where $n > -3$ is the spectral index of the initial power spectrum and $m_*$
stands for the large-mass cutoff.  In fact, 
the agreement between the power-law parts of Eq.~(\ref{P-S_MF}) and of the
found mass function demands  
$n \ra -3$. Therefore, we take
\begin{equation}
N(m) \approx N(1)\, m^{-2}
\exp\left[-\left(\frac{m}{m_*}\right)^{\!\e}\,\right],
\label{Nm}
\end{equation}
where $\e >0$ is to be fitted, as well as $m_*$.  The latter can be deduced
from the condition $M_1 = 1$; namely,
\begin{eqnarray*}
M_1 &\approx& \frac{N(1)}{N} 
\sum_{m=1}^{\infty} \frac{\exp[-(m/m_*)^\e]}{m} \\
&\approx&
-\frac{N(1)}{N} 
\int_{m=1}^{\infty} dm\,\frac{\exp[-(m/m_*)^\e]}{m} \approx 
\frac{N(1)}{N} \,\ln m_* \,.
\end{eqnarray*}
It is independent of $\e$, and coincides with the value given by
Eq.~(\ref{M1}) if we identify $m_{{\rm max}}$ there with $m_*$. This
identification is natural, because the exponential form (\ref{Nm}) is just one
way of introducing a mass cutoff that is more adequate than the sharp cutoff
used in Eq.~(\ref{M1}).  We can see why the above quoted values of $m_{{\rm
max}}$, below 1000, actually mark the end of the power laws. The new values of
$m_{{\rm max}}$ are obtained from expression (\ref{Nm}) by requiring
$N(m_{{\rm max}})=1$. Thus, this model raises the estimations of $m_{{\rm
max}}$, but the new values depend on $\e$.  For $\e =1$, $m_{{\rm max}}$ is
equal to 2849 (dark matter) or 2022 (gas). Naturally, better estimations are
obtained by taking smaller $\e$. In fact, the Press-Schechter mass function
must be substituted by a lognormal mass function \cite{I4}, in which the power
$(m/m_*)^\e$ becomes $[\ln(m/m_*)]^2$.

\subsection{Multifractal spectra and cosmic web structure}
\label{MFsp}

\begin{figure}
\centering{\includegraphics[width=7.5cm]{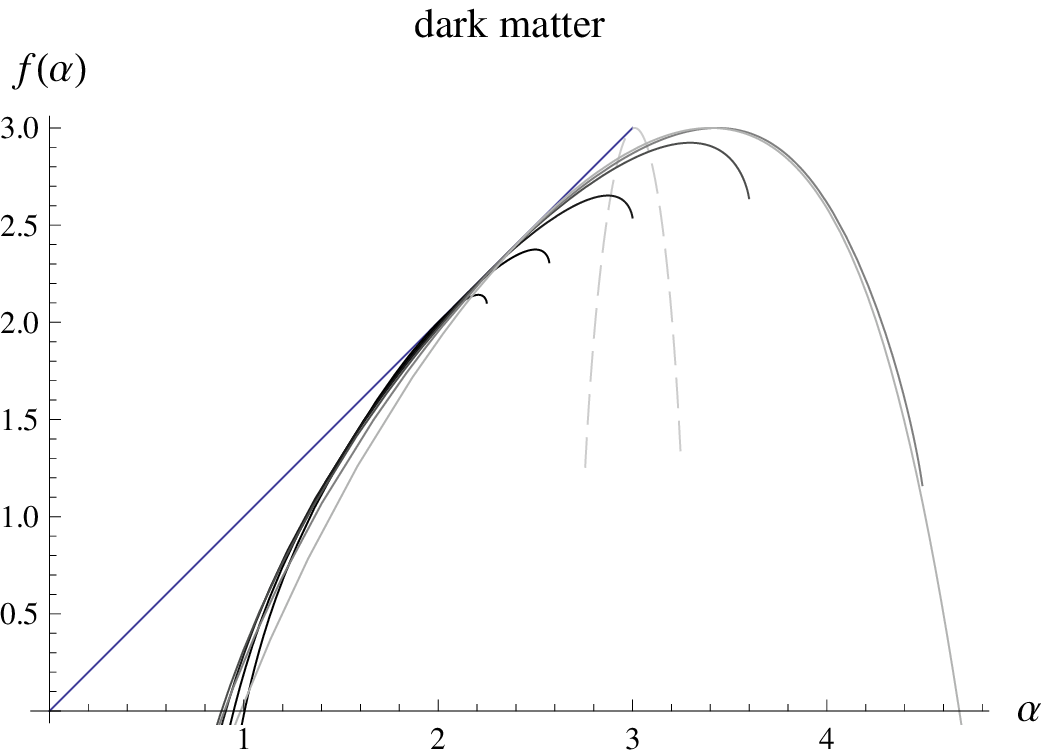}}
\centering{\includegraphics[width=7.5cm]{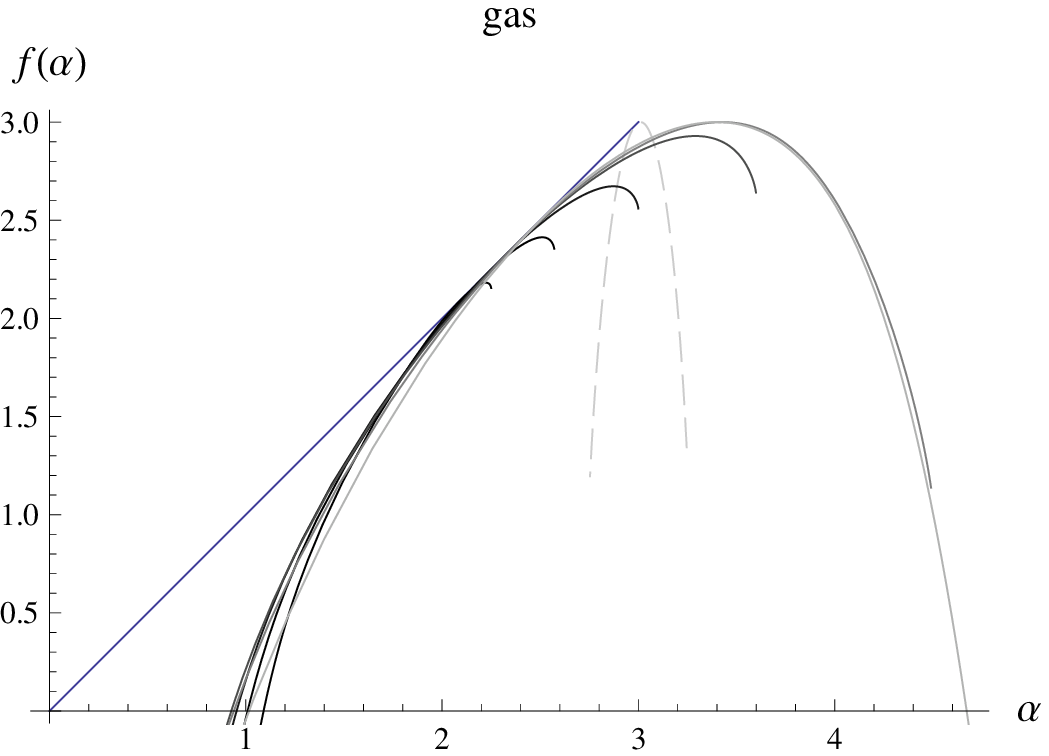}}
\caption{ Multifractal spectra at scales 
$l=2^{-12},2^{-11}, \ldots, 2^{-7}$, plotted with
solid lines in succesively lighter tones of grey.
The light dashed lines are the spectra of the homogeneous distributions at
$l=2^{-3}$.}  
\label{MFspec}
\end{figure}

The coarse multifractal spectrum is easily computed from the counts in cells,
through Eq.~(\ref{Mq}) and Eqs.~(\ref{aq}), (\ref{fa}) and (\ref{ctauq}).  We
plot in Fig.\ \ref{MFspec} the multifractal spectra of the dark-matter and gas
distributions at scales from $l=2^{-12}$ up to $l=2^{-7}$. We stop at this
scale because we already have the full spectrum, and on larger scales it
begins to show signs of a transition to homogeneity.  For comparison, we also
plot the spectra corresponding to the distributions at $l=2^{-3}$, which are
homogeneous [we have computed them using Eq.~(\ref{ctauq}) with $V_0 =1$].

The six multifractal spectra at successive scales coincide closely in their
respective ranges, except near $\a_{{\rm max}}$, and the spectra corresponding
to the dark matter are almost identical to the ones corresponding to the gas
(Fig.\ \ref{MFspec}).  In addition, they all are similar to the multifractal
spectra of the GIF2 simulation obtained in Ref.~\cite{I4}, although they span
a slightly larger range of local dimensions.  By increasing the reference
scale $V_0$ in Eq.~(\ref{ctauq}), we observe that the span of $\a$ at a given
scale shrinks, and thus we deduce that the slightly smaller spans in the GIF2
simulation are due to having set there no homogeneity scale ($V_0 =1$).  The
universal multifractal spectrum of cosmological distributions that all these
results suggest is typical of statistically self-similar multifractals.

The dimension of the mass concentrate in the spectra of Fig.\ \ref{MFspec}
slightly rises as the coarse-graining length grows; taking all the spectra
into account, we estimate $\a_1 \simeq 2.4$. This value agrees with the value
obtained from the GIF2 simulation. It is a remarkably high value, which makes
the mass concentrate relatively homogeneous.  It is interesting to consider
the meaning of this high dimension for a cosmic web structure. This type of
structure presumably possesses singularities of the three possible kinds,
namely, singular points, curves and surfaces, called nodes, filaments and
sheets, respectively.  At first sight, the high value of $D_1=\a_1$ may
suggest that the mass concentrates in the highest dimensional structures,
namely, sheets (Zel'dovich's ``pancakes''). However, self-similar
distributions of filaments or even of nodes can also reach fractal dimensions
higher than two.  Therefore, detailed morphological studies are necessary to
decide the relative weight of sheets, filaments and nodes in the cosmic web.

Morphological studies of multifractal distributions are by no means easy. In
fact, fractal dimensions do not reveal whether a distribution consists of
points, curves or surfaces. This information is given by the {\em topological
dimension}, whereas the fractal dimension informs about the clustering of
objects of given topological dimension.%
\footnote{The topological dimension is a topological invariant, unlike the
Hausdorff-Besicovitch (fractal) dimension. The topological dimension can be
defined in several equivalent ways and is always an integer: it is zero for
a point, one for a curve, two for a surface, etc.
The Hausdorff-Besicovitch dimension is bounded below by the topological
dimension. Actually, Mandelbrot \cite{Mandel} defines a fractal as a
set with Hausdorff-Besicovitch dimension strictly higher than its topological
dimension. Therefore, the degree of fractal clustering is measured by the
difference between both dimensions.}
Unfortunately, topological dimensions are very difficult to estimate from
finite samples of singular distributions.  One method of studying the topology
of a cosmic web finite sample has been devised by Sheth et al \cite{S4}. Their
method is based on a surface modelling algorithm (``SurfGen'').  Other methods
are described by van de Weygaert \& Schaap \cite{vW-Sch}, e.g., the method
based on the Delaunay tessellation field estimator.  Many morphological
studies of the cosmic web have focused on its voids, for the boundaries of
voids define the matter sheets (or vice versa); but there is no unique
definition of voids in finite samples.  Cosmic foams with self-similar
distributions of voids have relatively simple structures, with well defined
distributions of sheets, filaments and nodes.  Besides, the scaling of voids
is easily demonstrated in finite samples of these distributions.  However, the
cosmic web seems to be better described as a non-lacunar multifractal with
much more complex geometry \cite{I5}.%
\footnote{Note that this statement strictly applies to the full matter
distribution, whereas the cosmic web of galaxies could have a low lacunarity,
as discussed in Ref.~\cite{I5}.}

The dimension of the multifractal mass concentrate $\a_1 \simeq 2.4$ that we
find differs from standard determinations of the fractal dimension of the
galaxy distribution, which yield values close to two but usually smaller
\cite{Jones-RMP,Sylos-Pietro}.  However, this dimension is determined from the
two-point correlation function and, therefore, it corresponds to the
correlation dimension $\tau(2)=D_2$, which must be smaller than $\a_1=D_1$ (in
a multifractal). We determine $D_2$ in Sect.~\ref{frac_dist}.

Another interesting dimension is $D_0$, the box-counting dimension of the
distribution's support.  Since it coincides with the maximum of $f(\a)$, Fig.\
\ref{MFspec} shows that a reliable value of $D_0$ can only be obtained from
the scale $l=2^{-8}$ upwards. This value is 3, confirming the conclusion that
the cosmic web is a non-lacunar multifractal \cite{I5}.  Note that having
$D_0=3$ implies that 
the empty cells that appear in increasing numbers for $l \leq 2^{-8}$ are
actually empty because they belong to under-sampled zones.
Moreover, the high-$\a$ ends of 
the spectra at scales $l < 2^{-8}$ are given by scarcely occupied cells and,
naturally, deviate from the true spectrum (best represented at $l=2^{-7}$); in
particular, the maximum of $f(\a)$ is depressed, creating
the false impression of lacunarity. In fact, it
is necessary to suppress scarcely occupied cells to fully demonstrate scale
invariance, as we show next.

\subsection{Scaling of second order moments and correlation dimensions}
\label{frac_dist}

The superposition of the coarse spectra at $l=2^{-12}, \ldots, 2^{-7}$ in
their respective $\a$ ranges that is shown in Fig.~\ref{MFspec} constitutes a
proof of multifractality. However, the standard proof of scale invariance for
multifractals is based on the definition of $\tau$-exponents in
Eq.~(\ref{tauq}): scale invariance demands the scaling of $M_q$ in a range of
cell sizes that is sufficient to calculate a meaningful $\tau(q)$ and, hence,
$D_q$.  The exponent $\tau(q)$ is normally calculated by fitting the slope of
the plot of $\log M_q$ versus $\log l$.  We now follow this procedure.

First of all, we need to select the values of $q$ for which we calculate $M_q$
and also select the appropriate range of cell sizes. The available range of
$q$ is bound above by the condition that $f(\a) \geq 0$ (non-negative fractal
dimensions). This bound can be perceived in Fig.\ \ref{MFspec}, for the slopes
of the spectra do not become vertical at their left-hand ends, that is to say,
the respective values of $q=f'(\a)$ are bounded above.  The bound depends
somewhat on the particular spectrum and, in fact, becomes smaller as $l$
grows.  An examination of the numerical values of $q$ for the spectra plotted
in Fig.\ \ref{MFspec} reveals that the largest integer value of $q$ that is
common to all the spectra is $q=2$.  Note that the values of $f'(\a)$ differ
much more than the respective values of $f(\a)$.

The possible values of $q$ must also be bounded below: although most spectra
in Fig.\ \ref{MFspec} can be nominally extended to $q=f'(\a) \ra -\infty$,
this extension is inside their unreliable high-$\a$ ends. For example, it is
obvious from Fig.\ \ref{MFspec} that the spectrum at $l=2^{-12}$ (for either
dark matter or gas) does not represent well the mass concentrate,
corresponding to the point of contact with the diagonal.  Therefore, that
spectrum is not valid even down to $q=1$.  Consequently, when we combine the
upper and lower bounds, the only integer value allowed is $q=2$, so we must
restrict ourselves to examining the scaling of $M_2$ and calculating the
correlation dimension $D_2 = \tau(2)$.  Notice that this dimension is
of special interest, since it is the one that is usually measured in galaxy
surveys.

As regards the range of cell sizes in which to look for scaling, the natural
range lies between the homogeneity scale $l=2^{-4}$ and the discreteness scale
$l=2^{-10}$. However, we have seen above that the effects of under-sampling can
already be perceived at $l=2^{-8}$ and become more evident at $l=2^{-9}$.  On
the other hand, the cells that are well populated on scales $l \leq 2^{-8}$
are surely not affected by under-sampling, as proved by the superposition of
the spectra along their left-hand sides.  A sensible way to avoid the effects
of under-sampling in the computation of $M_2$ is to suppress for each $l$ the
scarcely occupied cells that contribute to the deviant piece of the
corresponding multifractal spectrum.  Thus, we set a lower cell-mass cutoff $m
= m_0 (l/l_0)^\a$, where $l_0=2^{-4}$, $m_0= N l_0 ^3 = 2^{18}$ and, for each
$l$, we choose the value of $\a$ that marks the beginning of the deviant
spectrum.  To be definite, we assume that the deviant pieces of the spectra
begin at their respective maxima (see Fig.\ \ref{MFspec}).  Thus, we can
proceed to $l<2^{-10}$, but we stop at $l = 2^{-12}$ because lower scales
present several problems: (i) the spectrum hardly represents the mass
concentrate; (ii) the value of $M_2$ becomes very sensitive to the precise
value of the $m$-cutoff; (iii) the gas distribution begins to noticeably
depart from the dark-matter distribution.

\begin{figure}
\centering{\includegraphics[width=7.5cm]{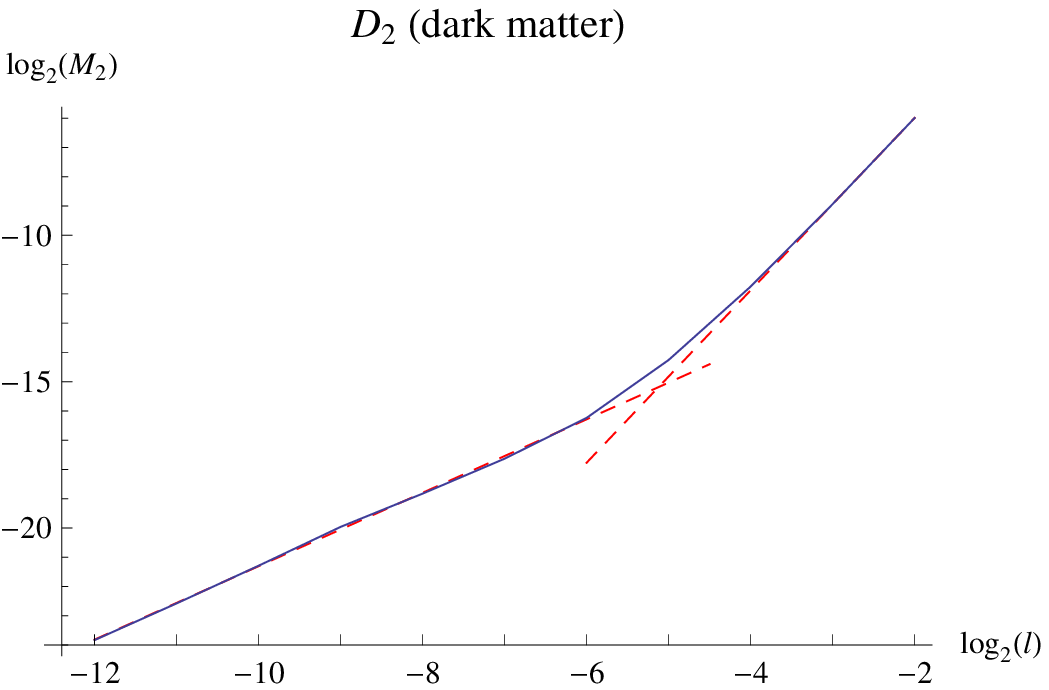}}
\centering{\includegraphics[width=7.5cm]{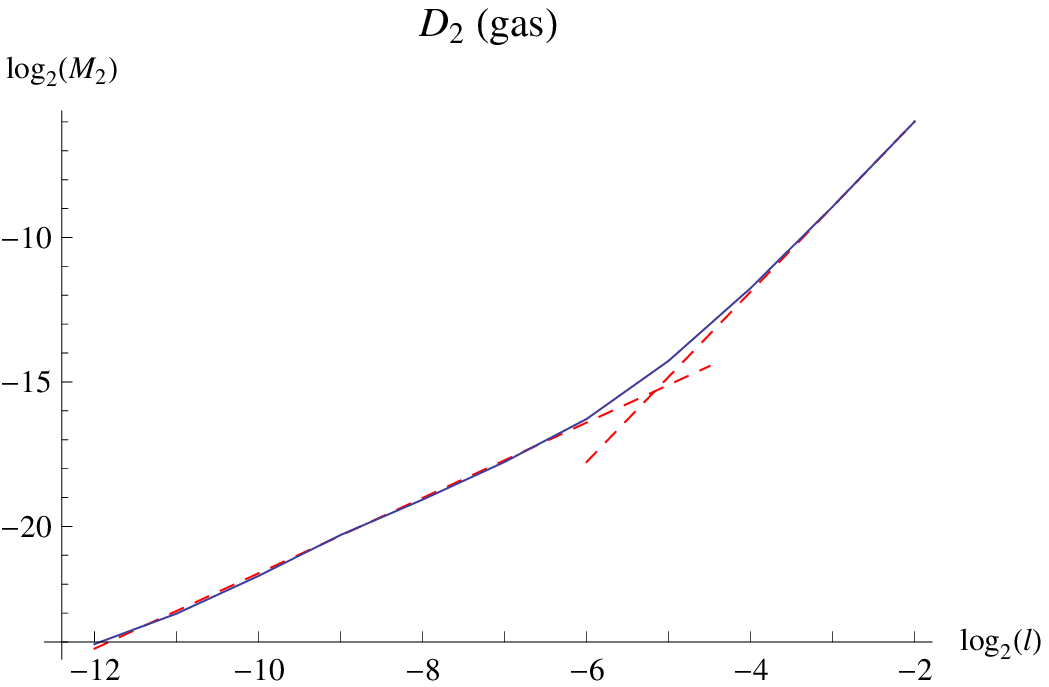}}
\caption{Log-log plots for the correlation dimension $D_2$ of
the distributions of dark-matter (left) and gas (right), 
showing the fractal scaling range and the transition to homogeneity.}
\label{D2-range}
\end{figure}

Therefore, we compute $M_2$ from the scale $l = 2^{-12}$ upwards, and actually
we do not stop at $l=2^{-4}$ but at $l=2^{-2}$, to study the full transition
to homogeneity.  The log-log plots of $M_2$ versus scale $l$ are displayed in
Fig.~\ref{D2-range}. The dashed straight lines correspond to the least-squares
fits. The two fits in the fractal ranges between $l = 2^{-12}$ and $l =
2^{-6}$ yield the following dimensions: (i) $D_2=1.255 \pm 0.012$ for the
dark-matter; (ii) $D_2=1.30 \pm 0.02$ for the gas.  The two fits in the
homogeneity ranges yield values of $D_2$ very close to three, of course.

In each plot, the scale at which the straight line of the fractal fit meets
the straight line of the homogeneity fit is a measure of the scale of
transition to homogeneity $l_0$. Thus, we deduce that $l_0 = 2^{-5}$,
approximately, for both the dark matter and the gas.  Notice that this measure
of the scale of homogeneity yields a smaller value than the one that we have
been using, $l_0 = 2^{-4}$. In fact, the transition to homogeneity is not very
sharp but takes place between $l = 2^{-6}$ and $l = 2^{-4}$, as
Fig.~\ref{D2-range} shows.

The value $D_2=1.30\pm 0.02$ for the gas is definitely smaller than the galaxy
correlation dimension $D_2= 2.0\pm 0.1$ obtained by Sylos Labini and
Pietronero \cite{Sylos-Pietro} but agrees with conventional values of $D_2$
\cite{Jones-RMP,Sylos-Pietro}. Sylos Labini and Pietronero's value stems from
their criticism of the treatment of finite size effects and, in particular,
from questioning the classical value of the scale of homogeneity $r_0 \simeq 5
\, h^{-1}\!$ Mpc. Indeed, they extend the scaling range of the correlation
function up to $30 \, h^{-1}\!$ Mpc and, consequently, $D_2$ grows. Other
authors also find $D_2 \simeq 2$, especially when they use long scale ranges
to compute it (see Table I in Ref.~\cite{Jones-RMP}). In our case, the end of
the scaling range at $l = 2^{-6}$ is quite clear, although we define as
homogeneity scale $l_0 = 2^{-5}$ (a fit up to this scale would hardly raise
$D_2$, anyway). The scale $l = 2^{-6}$ is $7.8 \, h^{-1}\!$ Mpc in physical
units.

To summarize the results of Sect.~\ref{MF}, the tests for scale invariance can
be considered successful, given the limitations imposed by the data.  Of
course, the scaling range necessary to affirm scale invariance is a matter of
opinion. A factor of $2^{-6}=64$ is reasonably good.  In addition to the
extent of a scaling range, one must also consider the quality of the
corresponding least-squares fit, namely, its standard error.  In this regard,
the fits for the dark-matter and gas distributions are both remarkably good.
We refrain from affirming that we have proved that these distributions are
(samples of) statistically self-similar multifractals, but we assert that
there is strong evidence of it.  Furthermore, there is good evidence that the
dark-matter and gas distributions are indistinguishable multifractals.  One
could object that the confidence intervals for the $D_2$ do not overlap and
that there are minute differences between the respective plots in Figs.\
\ref{P-S} and \ref{MFspec}.  To assess the statistical significance of the
numerical differences between the distributions of dark-matter or gas
particles, we carry out next a detailed study.

\section{Relation between the gas and dark-matter distributions}
\label{bias}

We see that the multifractal properties of the dark-matter and gas
distributions are very similar (along a considerable range of scales), which
suggests that the distributions could actually be identical.  In general, one
may ask if two finite samples of continuous distributions can come from the
same continuous distribution.  In particular, it is possible that the
differences between the distributions of gas and dark matter particles are
only due to statistical sample variance, while the continuous gas distribution
is unbiased with respect to the total mass distribution (dominated by the dark
matter).  We know that the gas dynamics is different from the collisionless
dark-matter dynamics, with the likely result of bias, but we need to ascertain
the existence of bias from the actual particle distributions by means of
statistical tests.

The first test that comes to one's mind is based on the cross-correlation
function of gas and dark-matter particles, in particular, the
cross-correlation coefficient, useful to measure the similarity of two
distributions. Indeed, this test confirms that both distributions are very
similar, as we show in sub-section \ref{cross}.  However, this test cannot
{\em prove} that the samples actually come from the same continuous
distribution. In fact, it is easy to see that there is no way to prove it and
we must satisfy ourselves with obtaining a probability of its being
true. Rather, assuming a Bayesian point of view, we can quantify the ``degree
of belief'' in the hypothesis that there is a common continuous distribution
(sub-section \ref{Bayes-sect}). The application of this method in
sub-sect.~\ref{appl_Bayes} allows us to confidently conclude that the gas
distribution is biased on nonlinear scales. Then, we study the nature of that
bias in sub-section \ref{G-DM_entropy}.

To compare the two distributions at several scales, we use counts in cells,
like in the multifractal analysis.  Thus, we assume that two independent
continuous distributions define the probabilities of the respective counts in
cells of given size.  In other words, we assume that the dark-matter and gas
distributions are both samples of respective multinomial distributions, each
one given by a set of probabilities defined in the cells.  The
cross-correlation can be easily expressed in terms of counts in cells.  The
Bayesian method seeks the probability (degree of belief) that two multinomial
samples come from the same multinomial distribution \ref{Bayes-sect}.

\subsection{Cross-correlations}
\label{cross}

Given a mass distribution coarse-grained with volume scale
$V$, its auto-correlation is measured by the second order cumulant
$${\bar\xi}_2 = \frac{1}{V^2} \int_V d^3x_1\, d^3x_2\, \xi_2(x_1,x_2),$$ 
where ${\xi}_2$ is the two-point correlation function of the fine grain
distribution.  In the nonlinear regime,
$${\bar\xi}_2 = \frac{\langle\r^2\rangle}{\langle\r\rangle^2}-1 \approx
\frac{\langle\r^2\rangle}{\langle\r\rangle^2} = \frac{\mu_2}{\mu_1^2} \gg 1.$$
We can define the {\em cross-correlation} coefficient of gas (g) and
dark-matter (m) at scale $V$ as
\begin{eqnarray*}
{c}_{\rm gm} &=& \frac{{\bar\xi}_{\rm gm}} {\left(\bar\xi_{\rm 2g}
\,\bar\xi_{\rm 2m}\right)^{1/2}} = \frac{\langle\r_{\rm g}\,\r_{\rm m}\rangle}
{\left(\langle\r_{\rm g}^2\rangle \,\langle\r_{\rm m}^2\rangle \right)^{1/2}}
= \frac{\sum_{i=1}^M n_{{\rm g}\,i}\, n_{{\rm m}\,i}} {\left(\sum_{i=1}^M
n_{{\rm g}\,i}^2\right)^{1/2}\left(\sum_{i=1}^M n_{{\rm
m}\,i}^2\right)^{1/2}}\,,
\end{eqnarray*}
where the last expression refers to counts in volume-$V$ cells and $M$ denotes
the number of these cells. The cross-correlation coefficient can be viewed as
the cosine of the angle formed by the two $M$-dimensional vectors $\{n_{{\rm
    g}\,i}\}$ and $\{n_{{\rm m}\,i}\}$.

\begin{figure}
\centering{\includegraphics[width=8cm]{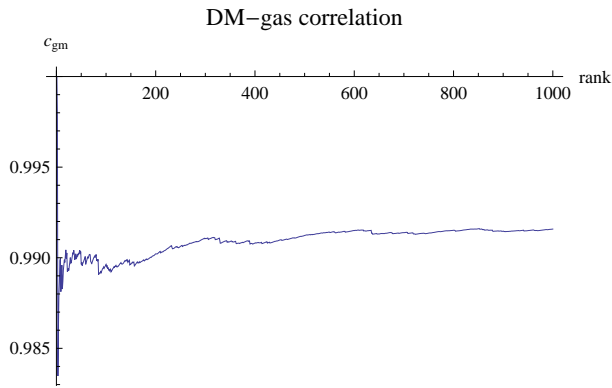}}
\caption{Cross-correlation coefficient of gas and dark matter in massive
halos, as a function of the number of halos, ranked in order of decreasing
mass.}
\label{correl}
\end{figure}

Given the cell counts $\{n_{{\rm g}\,i}\}$ and $\{n_{{\rm m}\,i}\}$, we can
compute ${c}_{\rm gm}$ at once, but we follow instead a more elaborate
procedure to discern the influence of the cell masses.  We first rank the
cells in order of decreasing {\em physical} mass, for physical mass determines
the importance of cells in regard to gravity.  Then, we compute the
cross-correlation coefficient of the ordered cells up to successive rank
values.  In Fig.~\ref{correl}, we plot the cross-correlation coefficient of
the gas and dark-matter in massive halos (taken from the master cell
distribution), computed in that way.  This coefficient is stably above 0.99,
that is to say, the correlation between both distributions is very strong.
Moreover, the cross-correlation coefficient increases with the coarse-graining
scale $l$. For example, it reaches 0.9999 at $l=2^{-5}$.  However, we have no
way of knowing how strong the correlation must be for allowing us to affirm
that both samples come from the same distribution.

\subsection{Bayesian comparison of multinomial distributions}
\label{Bayes-sect}
 
Bayes' theory of probability interprets the concept of probability as a
measure of a state of knowledge. Bayes' theorem tells us how to adjust
probabilities in regard to new evidence. It writes
$$P(H|E) = \frac{P(E|H)\,P(H)}{P(E)}\,,$$ where $H$ is a hypothesis with {\em
prior} probability $P(H)$, $E$ is an event that provides new evidence for $H$,
and $P(E|H)$ is the conditional probability of having $E$ if the hypothesis
$H$ happens to be true. $P(E)$ is the a priori probability of observing the
event $E$ under all possible hypotheses.  $P(H|E)$ adjusts $P(H)$ and is
called the posterior probability of $H$ given $E$.  Bayesian analysis is
routinely employed for model selection in many scientific areas.

In Bayes' theorem, the hypothesis $H$ can belong to a continuum of
possibilities. For example, if we are given the results of $N$ trials of a
binomial experiment, we can analyse the information gained from them on the
probability $p$ of ``success'' (``success'' is defined arbitrarily as one of
the two possible outcomes).  This probability is a number $0 < p < 1$ (while
the probability of ``failure'' is $1-p$). For a given value of $p$, the
probability of $n$ successes in $N$ trials is given by the binomial
distribution
$$P(N,n|p) = \binom{N}{n}\, p^n (1-p)^{N-n}.$$
Since $N$ and $n$ are given and $p$ is unknown, we can apply 
Bayes' theorem in the form
\begin{eqnarray*}
P(p|N,n) &=& \frac{P(N,n|p)\,P(p)}{\int_0^1 P(N,n|p)\,P(p)\,dp} 
= 
\frac{p^n (1-p)^{N-n}\,P(p)} {\int_0^1 p^n (1-p)^{N-n}\,P(p)\,dp}\,.
\end{eqnarray*}
It yields the probability of $p$ given the data in terms of the prior
probability of $p$. If no prior information about $p$ is available, we must
assume that $P(p)=1$ (according to the principle of insufficient reason).
Then, the posterior probability $P(p|N,n)$ is the {\em beta distribution} with
parameters $n+1$ and $N-n+1$. It is trivial to check that it reaches its
maximum at $p=n/N$ (mode value) and that its variance is proportional to $1/N$
(for fixed $n/N$ and large $N$).

This example is, in fact, relevant to our problem, namely, to estimating the
probability that the given gas and dark-matter samples belong to the same
distribution. If we choose one cell, with $n_{{\rm m}}$ dark-matter particles,
say, the probability that the mass fraction in that cell is $p_{{\rm m}}$ is
given by the beta distribution with parameters $n_{{\rm m}}+1$ and $N_{{\rm
m}}-n_{{\rm m}}+1$ ($N_{{\rm m}}$ being the total number of dark-matter
particles in the sample). Analogously, the probability of a gas mass fraction
$p_{{\rm g}}$ in that cell is given by the beta distribution with parameters
$n_{{\rm g}}+1$ and $N_{{\rm g}}-n_{{\rm g}}+1$. We can obtain the probability
of the difference $p_{{\rm m}}-p_{{\rm g}}$ by taking the product $P(p_{{\rm
m}}|n_{{\rm m}},N_{{\rm m}}) P(p_{{\rm g}}|n_{{\rm g}},N_{{\rm g}})$,
performing the change of the variables $p_{{\rm m}}$ and $p_{{\rm g}}$ to
$p_{{\rm m}}-p_{{\rm g}}$ and $(p_{{\rm m}}+p_{{\rm g}})/2$, and integrating
over the second variable (within the appropriate limits).  However, the
difference $p_{{\rm m}}-p_{{\rm g}}$ is a continuous variable and its
probability is a probability {\em density}; therefore, the probability that
$p_{{\rm m}}=p_{{\rm g}}$ vanishes. Nevertheless, we expect to get some
information from the value of the probability density at $p_{{\rm m}}=p_{{\rm
g}}$. Thus, we calculate
\begin{eqnarray}
\lefteqn{
\int_0^1 P(p|n_{{\rm m}},N_{{\rm m}})\, P(p|n_{{\rm g}},N_{{\rm g}})\, dp =}
\nonumber\\
& & \phantom{aaaaaaaa}
\frac{B(n_{{\rm m}}+n_{{\rm g}}+1, N_{{\rm m}}-n_{{\rm m}}+N_{{\rm g}}-
n_{{\rm g}}+1)}{B(n_{{\rm m}}+1, N_{{\rm m}}-n_{{\rm m}}+1)\,B(n_{{\rm g}}+1, 
N_{{\rm g}}-n_{{\rm g}}+1)}\,,
\label{b}
\end{eqnarray}
where ${B(x,y)=\G(x)\,\G(y)/\G(x+y)}$ is the Euler beta function.  The value
of the integral is enhanced when the maxima of $P(p|n_{{\rm m}},N_{{\rm m}})$
and $P(p|n_{{\rm g}},N_{{\rm g}})$ coincide, namely, when $n_{{\rm m}}/N_{{\rm
m}} = n_{{\rm g}}/N_{{\rm g}}$. For fixed $N_{{\rm m}} = N_{{\rm g}}$, the
function on the right-hand side of Eq.~(\ref{b}) is a symmetric function of
$\{n_{{\rm m}},n_{{\rm g}}\}$. Therefore, for a fixed value of $n_{{\rm
m}}+n_{{\rm g}}$, it is just a symmetric function of the difference $n_{{\rm
m}}-n_{{\rm g}}$ and has its maximum when $n_{{\rm m}}=n_{{\rm g}}$.

One could criticize the preceding approach for only focusing on the value of
the probability density at $p_{{\rm m}}=p_{{\rm g}}$, while values of $p_{{\rm
m}}-p_{{\rm g}}$ close to zero might also be relevant.  We can avoid the
problem of having to deal with a continuous probability by singling out the
value $p_{{\rm m}}=p_{{\rm g}}$ from the outset. Thus, we formulate a Bayesian
analysis with this hypothesis and the event $E=\{n_{{\rm m}},N_{{\rm
m}},n_{{\rm g}},N_{{\rm g}}\}$:
\begin{eqnarray*}
P(p_{{\rm m}}=p_{{\rm g}} | E)
= 
\frac{P(p_{{\rm m}}=p_{{\rm g}})\,P(E | p_{{\rm m}}=p_{{\rm g}})}
{P(p_{{\rm m}}=p_{{\rm g}})\,P(E | p_{{\rm m}}=p_{{\rm g}}) +
P(p_{{\rm m}} \neq p_{{\rm g}})\,P(E | p_{{\rm m}} \neq p_{{\rm g}})}\,.
\end{eqnarray*}
Here, $P(E | p_{{\rm m}} \neq p_{{\rm g}})$ is just the probability of $E$
given any values of $p_{{\rm m}}$ and $p_{{\rm g}}$, because the event
$p_{{\rm m}}=p_{{\rm g}}$ has probability zero; namely,
\begin{eqnarray*}
P(E | p_{{\rm m}} \neq p_{{\rm g}}) = 
\binom{N_{{\rm m}}}{n_{{\rm m}}} \binom{N_{{\rm g}}}{n_{{\rm g}}} 
\int_0^1 dp_{{\rm m}} 
\int_0^1 dp_{{\rm g}} \,
p_{{\rm m}}^{n_{{\rm m}}} (1-p_{{\rm m}})^{N_{{\rm m}}-n_{{\rm m}}}\,
p_{{\rm g}}^{n_{{\rm g}}} (1-p_{{\rm g}})^{N_{{\rm g}}-n_{{\rm g}}}.
\end{eqnarray*}
On the other hand,
\begin{eqnarray*}
P(E | p_{{\rm m}} = p_{{\rm g}}) &=&
\binom{N_{{\rm m}}}{n_{{\rm m}}} \binom{N_{{\rm g}}}{n_{{\rm g}}} 
\int_0^1 dp\,
p^{n_{{\rm m}}+n_{{\rm g}}} (1-p)^{N_{{\rm m}}-n_{{\rm m}} + N_{{\rm
      g}}-n_{{\rm g}}}.
\end{eqnarray*}
Computing the integrals and substituting, we obtain
\begin{eqnarray*}
P(p_{{\rm m}}=p_{{\rm g}} | E) = \frac{P(p_{{\rm m}}=p_{{\rm g}})\, b(n_{{\rm
m}},N_{{\rm m}};n_{{\rm g}},N_{{\rm g}})} {P(p_{{\rm m}}=p_{{\rm g}})\,
b(n_{{\rm m}},N_{{\rm m}};n_{{\rm g}},N_{{\rm g}}) + P(p_{{\rm m}} \neq
p_{{\rm g}})\, }\,,
\end{eqnarray*}
where
\begin{eqnarray}
b(n_{{\rm m}},N_{{\rm m}};n_{{\rm g}},N_{{\rm g}}) &=& 
\frac{P(E | p_{{\rm m}}=p_{{\rm g}})}{P(E | p_{{\rm m}} \neq p_{{\rm g}})} 
\label{Bf1}\\  &=&
\frac{B(n_{{\rm m}}+n_{{\rm g}}+1, N_{{\rm m}}-n_{{\rm m}}+N_{{\rm g}}-
n_{{\rm g}}+1)}
{B(n_{{\rm m}}+1, N_{{\rm m}}-n_{{\rm m}}+1)\,B(n_{{\rm g}}+1, 
N_{{\rm g}}-n_{{\rm g}}+1)}\,.
\label{Bf2}
\end{eqnarray}
This function of $\{n_{{\rm m}},N_{{\rm m}},n_{{\rm g}},N_{{\rm g}}\}$
coincides with the value of the probability density of $p_{{\rm m}}-p_{{\rm
g}}$ at 0 given by Eq.~(\ref{b}). Therefore, this approach is consistent with
the preceding one: if $b(n_{{\rm m}},N_{{\rm m}};n_{{\rm g}},N_{{\rm g}})$ is
large, then $P(p_{{\rm m}}=p_{{\rm g}} | E)$ tends to one, independently of
the prior probability $P(p_{{\rm m}}=p_{{\rm g}})$. However, we have no way of
estimating this prior probability.

The assignment of prior probabilities is a usual problem in Bayesian
analyses, to the extent that Bayes' theory of probability has been deemed
subjective. However, there is no subjectivity if we indeed understand
Bayes' theory as a way of adjusting probabilities in regard to new evidence. 
The {\em Bayes factor} defined in Eq.~(\ref{Bf1}) is such that
\begin{eqnarray*}
\log 
\frac{P(p_{{\rm m}}=p_{{\rm g}} | E)}{P(p_{{\rm m}} \neq p_{{\rm g}} | E)} =
\log b(n_{{\rm m}},N_{{\rm m}};n_{{\rm g}},N_{{\rm g}}) +
\log
\frac{P(p_{{\rm m}}=p_{{\rm g}})}{P(p_{{\rm m}} \neq p_{{\rm g}})}\,.
\end{eqnarray*} 
Hence, we can endow this equation with an information theory meaning: the
prior information about the odds of our hypothesis is updated by the
information $\log b$ provided by the event $E=\{n_{{\rm m}},N_{{\rm
    m}},n_{{\rm g}},N_{{\rm g}}\}$.  The prior information is null if
$P(p_{{\rm m}}=p_{{\rm g}})=P(p_{{\rm m}} \neq p_{{\rm g}})$, but the
information provided by the event is independent of any prior probabilities.
The information provided by $E$ is positive or negative according to whether
the Bayes factor is larger or smaller than one.  The addition of informations
is independent of the (common) base of the logarithms, but it is convenient to
use base two and measure the information in bits. If the Bayes factor is
larger than one half and smaller than two, the information provided by $E$ is
smaller than one bit and can hardly be considered significant. For example,
with $N_{{\rm m}} = N_{{\rm g}}=200$, $\log_2 b(100,200;100,200) = 3.00$ bits,
$\log_2 b(100,200;80,200) = 0.11$ bits, and $\log_2 b(100,200;70,200) = -3.61$
bits, and only the first case or the last case provide evidence for or against
$p_{{\rm m}}=p_{{\rm g}}$, respectively.

Since we actually divide the sample into many cells, we need to generalize the
above method of comparing binomial distributions to the case of multinomial
distributions. This generalization is straightforward, except that we now have
to take care of normalizing the $P(E | \cdot)$ such that $\sum_E P(E | \cdot)
=1$.  The resulting Bayes factor is
\begin{eqnarray*}
\lefteqn{ b(n_{{\rm m}\,1}, \ldots, n_{{\rm m}\,k};n_{{\rm g}\,1}, \ldots, 
n_{{\rm g}\,k}) =}\\
& & \phantom{aaaaaaaa}
\frac{B(n_{{\rm m}\,1}+n_{{\rm g}\,1}+1, \ldots , n_{{\rm m}\,k} +
n_{{\rm g}\,k}+1)}{B(n_{{\rm m}\,1}+1, \ldots, n_{{\rm m}\,k}+1)\,B(n_{{\rm
g}\,1}+1, \ldots, n_{{\rm g}\,k}+1)\,(k-1)!}\,,
\end{eqnarray*}
where $\{n_{{\rm m}\,i}\}_{i=1}^k$ and $\{n_{{\rm g}\,i}\}_{i=1}^k$ are the
vectors denoting the numbers of dark-matter and gas particles, respectively,
in the $k$ cells, and $B(x_{1}, \ldots, x_{k}) = \G(x_{1}) \cdots
\G(x_{k})/\G(x_{1} + \cdots + x_{k})$ is the generalized Euler beta function.
We can write this Bayes factor as follows:
\begin{eqnarray}
\lefteqn{ b(n_{{\rm m}\,1}, \ldots, n_{{\rm m}\,k};n_{{\rm g}\,1}, \ldots, 
n_{{\rm g}\,k}) = }\nonumber\\ 
& & \phantom{aaaaaaaaaaa}
\binom{n_{{\rm m}\,1}+n_{{\rm g}\,1}}{n_{{\rm m}\,1}}
\cdots
\binom{n_{{\rm m}\,k}+n_{{\rm g}\,k}}{n_{{\rm m}\,k}} 
\frac{(N_{{\rm m}}+k-1)!\,(N_{{\rm g}}+k-1)!}{(N_{{\rm m}}+N_{{\rm
      g}}+k-1)!\,(k-1)!}\,, 
\label{entBf}
\end{eqnarray}
where $N_{{\rm m}} = n_{{\rm m}\,1}+ \cdots + n_{{\rm m}\,k}$ and $N_{{\rm g}}
= n_{{\rm g}\,1}+ \cdots + n_{{\rm g}\,k}$ are the total numbers of
dark-matter and gas particles, respectively (which are equal, in our case).
The latter form has the advantage of being the product of $k$ binomial
numbers, one per cell, times an overall factor. Each
binomial number expresses the number of ways of dividing the total number of
particles in the corresponding cell between the respective numbers of 
gas and dark-matter particles.
We can associate the (base-two) logarithm of that binomial number with a
``cell entropy''. This entropy is maximal when the numbers of dark-matter and
gas particles in the cell are equal and vanishes when there are no particles
of one type in the cell.

Let us take $N_{{\rm m}} = N_{{\rm g}} = N$.  To compute the Bayes factor, we
follow an analogous procedure to the one employed to compute the
cross-correlation coefficient ${c}_{\rm gm}$.  Since the above-described
Bayesian analysis is valid for any multinomial distribution or, in other
words, the cells are of logical rather than physical nature, we can group
several physical cells into one.  In particular, we can group the less
significant cells, namely, the ones with small numbers of particles. A
systematic procedure for grouping the cells consists in ordering them by
decreasing total number of particles and separating the most populated ones to
take them first into account. Thus, we take the first rank cell and compare it
against the remainder, using the binomial Bayes factor.  The evidence for or
against $p_{{\rm m}}=p_{{\rm g}}$ cannot be considered definitive yet. Then,
we proceed to calculate the Bayes information of the two more populated cells
plus the ``cell'' with the remainder, and so onwards. If a definite trend is
soon established, that is to say, if the absolute value of the Bayes
information grows steadily, we consider it as a solid evidence for or against
the hypothesis, according to the sign of $\log_2 b$.

\subsection{Bayesian analysis of the distributions at several scales}
\label{appl_Bayes}

Here, we apply the above-explained procedure of systematic multinomial
Bayesian analysis to some relevant cell distributions. We prefer to rank the
cells again in order of decreasing {\em physical} mass, as in
Sect.~\ref{cross}, rather than in order of decreasing total number of
particles.

\begin{figure}
\centering{\includegraphics[width=7.5cm]{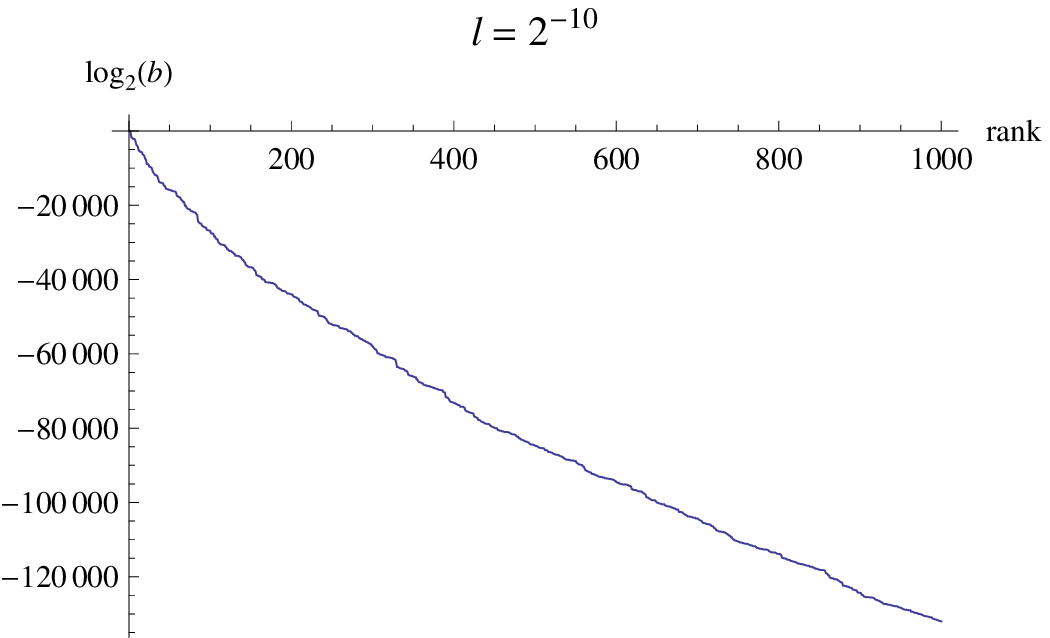}
\includegraphics[width=7.5cm]{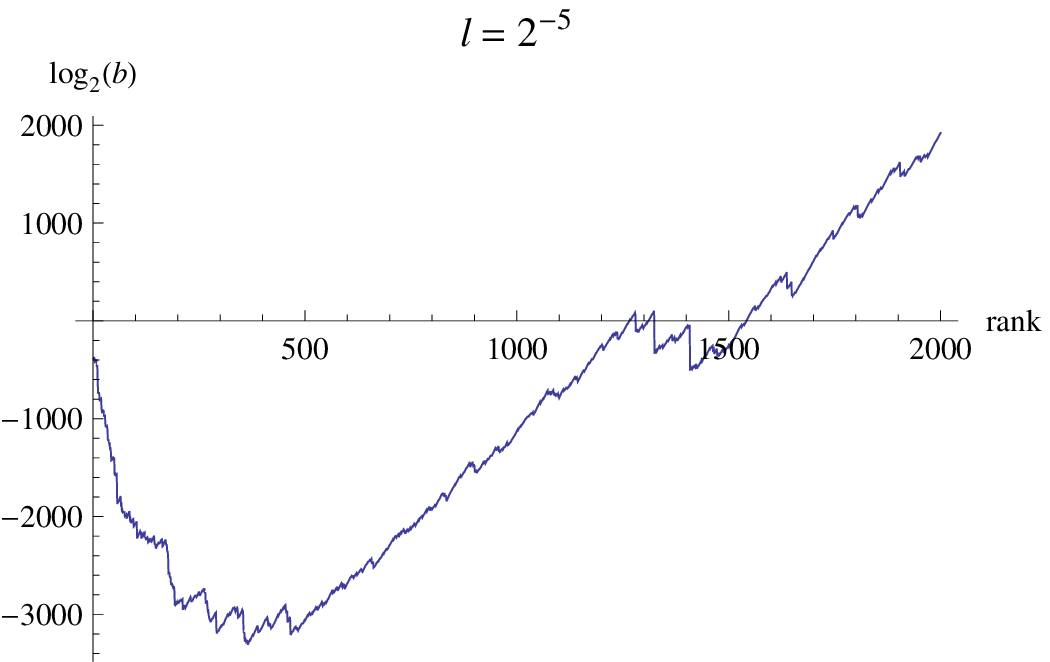}}
\caption{Bayesian evidence (in bits) for the equality of distributions
  ($p_{{\rm m}}=p_{{\rm g}}$) derived from massive cells at $l=2^{-10}$
  (halos) and at $l=2^{-5}$ (transition to homogeneity).}
\label{Bayes}
\end{figure}

We calculate the Bayes information $\log_2 b$ (in bits) for the hypothesis
$p_{{\rm m}}=p_{{\rm g}}$, considering a growing number of the most massive
cells.  The result is plotted in Fig.~\ref{Bayes}, for the two most relevant
scales: $l=N^{-1/3}=2^{-10}$ (corresponding to the master cell distributions)
and $l=2^{-5}$ (the scale of transition to homogeneity).  In the first case,
we see that the 1000 most massive halos already show that the evidence against
the hypothesis is overwhelming: note that $\log_2 b$ reaches $-130$ Kbits and
keeps its downward tendency.  The evidence in the second case is mixed: it is
increasingly negative up to the 500th rank, reaching $-3$ Kbits, but there it
starts growing and becomes positive from the 1550th rank onwards (the 1550th
cell contains $84\hspace{1pt}678$ dark matter particles and
$83\hspace{1pt}885$ gas particles).  Considering that the total number of
cells is $l^{-3}=2^{15}=32\hspace{1pt}768$ and that the corresponding total
Bayes information is 185.4 Kbits, we could say that the evidence of the
hypothesis $p_{{\rm m}}=p_{{\rm g}}$ is sufficient. However, the most massive
cells clearly distinguish both distributions.

Proceeding to larger scales, namely, to $l=2^{-4}$, the above pattern
holds. At $l=2^{-4}$, the Bayes information has some small fluctuations about
zero in the first ranks, staying above $-97$ bits, and then it definitely
grows, reaching a total of 28.8 Kbits. In this case, the evidence for $p_{{\rm
m}}=p_{{\rm g}}$ is solid. Of course, the evidence for $p_{{\rm m}}=p_{{\rm
g}}$ is stronger at larger $l$.

\begin{figure}
\centering{\includegraphics[width=8cm]{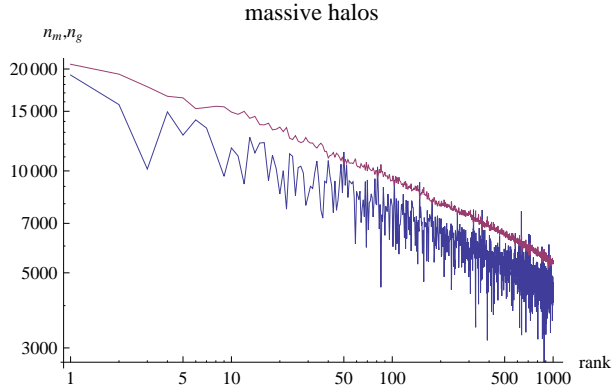}}
\caption{Particle counts of dark-matter (upper line) and gas (lower line) 
in halos ranked in order of decreasing mass.}
\label{Z_DM-G}
\end{figure}

Regarding the origin of the difference between $p_{{\rm m}}$ and $p_{{\rm g}}$
on small scales, let us focus on the master cell distributions.  An inspection
of the dark-matter and gas particle counts in massive halos reveals that these
consistently have fewer gas particles than dark-matter particles.  The smaller
average number of gas particles is clearly observed in the respective log-log
plots of counts in cells ranked by total physical mass, which are shown in
Fig.~\ref{Z_DM-G}.  We observe in the figure that both distributions
approximately follow linear log-log laws (sort of Zipf's laws), with common
slope, but the line that corresponds to the dark-matter particles is
definitely above. In other words, the massive halos concentrate less gas,
although the number of gas particles decreases according to the same pattern
that the number of dark-matter particles. One can also notice that there are
more fluctuations in the number of gas particles, due to their smaller
physical mass.

It is useful to express the differences between dark-matter and gas particle
counts in terms of the cell entropies introduced in Sect.~\ref{Bayes-sect}
after Eq.~(\ref{entBf}), as we do next.

\subsection{Entropic difference between the gas and dark-matter distributions}
\label{G-DM_entropy}

In the expression (\ref{entBf}) of the Bayes factor, we can consider that the
$k$ cells consist of a small number of massive cells $h = k-1$ (in order of
decreasing mass) and a $k$-th ``cell'' containing the remaining particles.
Furthermore, we assume that the $h$ massive cells contain together a total
number of particles that is small in comparison with the total number of
particles.  Recalling that $N_{{\rm m}} = N_{{\rm g}} = N= 2^{30} \gg 1$, we
can make a suitable approximation of the Bayes information $\log_2 b$.
Indeed, under the given conditions, the largest contributions to the Bayes
information come from the cell with the remaining particles and from the
overall factor in Eq.~(\ref{entBf}); namely,
\begin{eqnarray*}
\log_2 
\binom{2N-\sum_{i=1}^h (n_{{\rm m}\,i}+n_{{\rm g}\,i})}%
{N - \sum_{i=1}^h n_{{\rm g}\,i}} = 
2N - \sum_{i=1}^h (n_{{\rm m}\,i}+n_{{\rm g}\,i}) - \frac{\log_2(\pi N)}{2} +
{\Mfunction{O}(N^{-1})}
\end{eqnarray*}
and
\begin{eqnarray*}
\log_2 
\frac{(N+h)!\,^2}{(2N+h)!\, h!} = 
-2N + h\,\log_2\frac{N}{2} + \frac{\log_2(\pi N)}{2} +
{\Mfunction{O}(N^{-1})}- \log_2 h!\,,
\end{eqnarray*}
where we have used Stirling's approximation.  Note that both contributions
have a first term proportional to $N$, but these large terms cancel one
another. Therefore,
\begin{eqnarray}
\log_2 b &=& \sum_{i=1}^h \left[
\log_2\binom{n_{{\rm m}\,i}+n_{{\rm g}\,i}}{n_{{\rm g}\,i}} - 
(n_{{\rm m}\,i}+n_{{\rm g}\,i}) \right] + \nonumber\\
&&
h\,\log_2\frac{N}{2} - \log_2 h! + {\Mfunction{O}(N^{-1})}\,,
\label{Bayes_info}
\end{eqnarray}
which only grows logarithmically with $N$.  This Bayes information is a sum of
individual cell contributions plus a global contribution.  Each cell
contribution is negative, because the cell entropy is bounded above by the
number of particles in the cell, as is easily proved. If each massive cell
contribution is larger in absolute value than $\log_2(N/2)=29$ bits, on
average, the total information due to the $h$ massive cells plus the remainder
is negative.

In particular, each massive halo contributes, on average, more than
$\log_2(N/2)=29$ bits (in absolute value). This is the reason why the total
Bayes information of a considerable number of massive halos is negative (as
shown in Fig.~\ref{Bayes}).  For example, the contribution of the most massive
halo, with $n_{{\rm g}} = 19\hspace{1pt}200$ and $n_{{\rm m}} =
20\hspace{1pt}658$, is $\log_2\binom{39\hspace{1pt}858}{19\hspace{1pt}200} -
39\hspace{1pt}858 = -38.5$ bits, larger in absolute value than 29 bits.

The contribution of a massive cell to the Bayes information can be expressed
in a more familiar form by using again Stirling's approximation. When
$n_{{\rm m}}, n_{{\rm g}} \gg 1$, the cell entropy can be written as
\begin{eqnarray}
\log_2\binom{n_{{\rm m}}+n_{{\rm g}}}{n_{{\rm g}}} 
&\approx&  
(n_{{\rm m}}+n_{{\rm g}}) \log_2(n_{{\rm m}}+n_{{\rm g}})
- {n_{{\rm g}}} \log_2 {n_{{\rm g}}}
- {n_{{\rm m}}} \log_2 {n_{{\rm m}}} \label{mix}\\
&=& -(n_{{\rm m}}+n_{{\rm g}}) \left[{x_{{\rm g}}}
\log_2 {x_{{\rm g}}} + (1-{x_{{\rm g}}}) \log_2 (1-{x_{{\rm g}}})
\right], 
\end{eqnarray}
where we have introduced the fraction of gas particles
$$
{x_{{\rm g}}} = \frac{n_{{\rm g}}}{n_{{\rm m}}+n_{{\rm g}}}\,
$$ 
(the cell entropy has an analogous expression in terms of the fraction of
dark-matter particles).  In those forms, the cell entropy can be identified
with the familiar {\em entropy of mixing} \cite{Reif}.  Given $x_{{\rm g}}$,
the cell entropy of mixing is proportional to the total number of particles in
the cell; and so is the cell's contribution to the Bayes information, the
proportionality constant being the entropy of mixing per particle minus one.
The maximum entropy of mixing per particle is one bit and it corresponds to
the most mixed distribution, with $x_{{\rm g}} = 1/2$.  Naturally, a fully
mixed cell makes a vanishing contribution to the Bayes information.

Regarding the master cell distributions, we observe in Fig.~\ref{Z_DM-G} that
the ratio $n_{{\rm g}\,i}/n_{{\rm m}\,i}$ for massive halos is almost constant
on average; in fact, $n_{{\rm g}\,i}/n_{{\rm m}\,i} \simeq 0.81$. Hence,
$x_{{\rm g}\,i} \simeq 0.45$, and the entropy of mixing per particle is almost
constant and equal to
$$-{x_{{\rm g}}} \log_2 {x_{{\rm g}}} - (1-{x_{{\rm g}}}) \log_2
(1-{x_{{\rm g}}}) \simeq 0.992.$$
Therefore, each halo contribution is roughly proportional to the total number
of particles in it, with a common proportionality constant, namely, $0.992 - 1
= -0.008$. This yields about $-250$ bits for the contribution per halo in
Eq.~(\ref{Bayes_info}).  Thus, the absolute value of every massive halo
contribution is larger than 29 bits, making the total Bayes information in
Eq.~(\ref{Bayes_info}) negative and regularly decreasing with the number of
halos, as displayed in Fig.~\ref{Bayes} (left).  However, the value of the
entropy per particle is very close to one, telling us that the distributions
are very mixed, even though not completely mixed.

Note that a constant ratio $n_{{\rm g}\,i}/n_{{\rm m}\,i}$ for all the cells
would be in contradiction with $N_{{\rm g}} = N_{{\rm m}}$.  Thus, the ratio
$n_{{\rm g}\,i}/n_{{\rm m}\,i} \simeq 0.81$, for example, must grow
eventually, as $i$ runs over scarcely occupied cells. Even assuming that the
ratio $n_{{\rm g}\,i}/n_{{\rm m}\,i}$ stays almost constant as the cell mass
diminishes, the contribution per cell to the Bayes information is proportional
to the total number of particles in it and, therefore, it must eventually
become smaller than 29 bits (in absolute value). For one reason or another,
the initial downward trend of the Bayes information must cease and turn
upwards.  This turn is observed in the plot for $l = 2^{-5}$ in
Fig.~\ref{Bayes}.

\subsubsection{Connection with thermodynamics}

In thermodynamics, the entropy of mixing is, of course, only one part of the
total entropy.  When other thermodynamic parameters are equal, the entropy of
mixing determines the equilibrium configuration to be the most mixed
distribution.  In our context, the gas and the dark matter are not comparable
thermodynamically because, in principle, cold dark matter does not have
temperature or pressure. However, CDM particles have velocity dispersion, such
that one can assign it a temperature and, hence, a thermodynamic entropy
(independent of the properties of the gas).  This is the dark-matter entropy
considered by Faltenbacher et al \cite{Falten} in their study of the entropy
of gas and dark-matter clusters from the Mare-Nostrum universe.  Once the dark
matter is assigned thermal states, it is legitimate to compare them with the
thermal states of the gas.

In a mixture of ideal gases, the chemical
potential of each gas can be expressed as 
$$
\mu = -T \log_2 \frac{\z(T)}{n}
$$
\cite{Reif}, where $n$ is the number density, $\z(T)$ is an increasing
function of $T$ characteristic of each gas, and we use units consistent with
measuring the entropy in bits.  The function $\z(T)$ is calculated from the
possible states of the gas particles (translational and internal states); for
a monoatomic gas, $\z(T) \propto T^{3/2}$.  The condition of ``chemical''
equilibrium of gas and dark matter is
$$
\frac{\mu_{{\rm g}}}{T_{{\rm g}}} = \frac{\mu_{{\rm m}}}{T_{{\rm m}}}\,,
$$
which allows for ${T_{{\rm g}}} \neq {T_{{\rm m}}}$. In fact, 
chemical equilibrium implies
$$
\frac{\z_{{\rm g}}(T)}{n_{{\rm g}}} = \frac{\z_{{\rm m}}(T)}{n_{{\rm m}}} 
\,\Ra \,
\frac{\z_{{\rm g}}(T)}{\z_{{\rm m}}(T)} =
\frac{n_{{\rm g}}}{n_{{\rm m}}},  
$$
and therefore different temperatures for different densities.  We have seen
above that ${n_{{\rm g}}}/{n_{{\rm m}}} \simeq 0.81$ for massive halos.
Hence, assuming that both $\z_{{\rm g}}$ and $\z_{{\rm m}}$ correspond to
monoatomic gases, we deduce that $T_{{\rm g}}/T_{{\rm m}} \simeq 0.87$.

The conclusion that the dark matter temperature is higher than the gas
temperature in massive halos may seem counterintuitive. But note that it
relies on the assumption of independent local thermodynamical equilibria of
dark matter and gas at different but well-defined temperatures, with the local
temperature of dark matter given by its local velocity dispersion.  This
assumption should imply that the dark matter also has pressure and, therefore,
its dynamics should be governed by similar equations to the ones that govern
the gas dynamics. However, the effects of dark-matter pressure are not
considered in the Mare-Nostrum or other $N$-body cosmological simulations.

\section{Entropic comparison of distributions}
\label{entropies}

In the comparison of the gas and dark-matter distributions, we have found it
useful to introduce a cell entropy, recognizable as the entropy of mixing.  In
general, the Boltzmann-Gibbs-Shannon (BGS) entropy of a discrete probability
distribution $\{p_i\}_{i=1}^M$ is defined as
\begin{equation}
S(\{p_i\}) = -\sum_{i=1}^M p_i \log_2 p_i\,,
\label{S}
\end{equation}
and it represents the uncertainty or lack of information of the result of an
experiment with that probability distribution.  Note that we are using now
``discrete'' in the normal sense of the word in probability theory, namely,
meaning that there is a list of possible events, as opposed to the continuum
of possible events in a {\em continuous} distribution; but the probabilities
$p_i$ are continuous variables.  The entropy has some desirable properties,
such as the bounds $0 \leq S(\{p_i\}) \leq \log_2 M$, and the property of
additivity, in particular, additivity for independent sets of events
\cite{Renyi}.  This property and the bounds are shared by a uni-parametric
class of functions, the R\'enyi entropies
\begin{equation}
S_q(\{p_i\}) = \frac{\log_2 (\sum_{i=1}^M {p_i}^q)}{1-q}, \quad 
q \neq 1\,.
\label{Sq}
\end{equation}
The value of $S_1$ is obtained as the limit $q \to 1$ and it coincides with
the standard BGS entropy defined by Eq.~(\ref{S}).

We can apply the definition of entropy to a {\em discrete} distribution of $N$
particles in $M$ cells, with occupation numbers $\{n_i\}_{i=1}^M$ (counts in
cells) and hence expected probability distribution $\{p_i = n_i/N\}_{i=1}^M$.
The entropy measures the uncertainty of the cell in which an arbitrary
particle is located (or a group of $q$ particles, in the case of $S_q$ with $q
\in \mathbb{N}$).  In particular, we can interpret Eq.~(\ref{S}) as follows.
According to Boltzmann, one should weight a macroscopical state, given by a
set of occupation numbers, with the number of microscopical states compatible
with it (the Boltzmann weight). Then, the entropy is the logarithm of this
weight.  Since the number of states compatible with the occupation numbers
$\{n_i\}_{i=1}^M$ is given by the corresponding multinomial number, the
entropy is given by the logarithm of that multinomial number, namely,
$$
\log_2\binom{N}{n_{1} \cdots n_{M}} \approx  - N \sum_{i=1}^M p_i \log_2 p_i 
\,,
$$
where we have assumed that $n_i \gg 1$, equivalent to neglecting the effect of
particle discreteness.  The entropy per particle $S(\{p_i\})$ is positive and
bounded above by $\log_2 M$.  If the distribution is uniform, the bound is
reached; in particular, the bound is $\log_2 M = -\log_2 V$.  Then, the
distribution contains the largest uncertainty or, equivalently, the smallest
information. Moreover, all the R\'enyi entropies reach the same bound.

Naturally, it is important to know the behaviour of the entropies in the
continuum limit of the discrete distribution $\{p_i\}_{i=1}^M$, as the cell
size $V \to 0$ and $M \ra \infty$ (for the distribution of $N$ particles in
$M$ cells, one must let $N \ra \infty$ before $M \ra \infty$).  Not
surprisingly, the entropies diverge in the continuum limit: one needs an
infinite amount of information to locate a point in a continuum.  R\'enyi
\cite{Renyi} describes the growth of the $S_q$ as the distribution becomes
continuous in terms of {\em dimensions}; namely, he defines for the continuous
distribution the dimensions
$$
D_q = \lim_{V \to 0} \frac{3\,S_q(\{p_i\})}{-\log_2 V},
$$ 
assuming that the limit exits.  These R\'enyi dimensions are standard in
multifractal analysis; they have been already introduced in Sect.~\ref{anal},
Eq.~(\ref{Dq}), and used in subsequent sections.  The most important R\'enyi
dimension is $D_1$, which is defined by the divergence of the standard BGS
entropy and is the dimension of the set of singularities where the probability
concentrates.  Since the full set of R\'enyi dimensions characterizes the
information content of the distribution in the continuum limit, we deduce that
all the continuous distributions with the same spectrum of R\'enyi dimensions
appear equivalent in regard to their information content.  In particular,
every continuous distribution with $D_q = 3, \; q \in \mathbb{R},$ appears
equivalent to a homogeneous and uniform distribution, in which the R\'enyi
entropies reach their upper bound (note that $D_q = 3$ is the upper bound to
the R\'enyi dimensions).  Indeed, only part of the information contained in a
continuous distribution is preserved in its R\'enyi dimensions.

One can further define the information content of a continuous distribution
in terms of its probability density \cite{Renyi}, if this density is well
defined.  However, we are studying distributions with singularities.  In a
singular distribution, the singularities must be confined to a set of zero
volume, but they can be crucial for determining the distribution (for example,
consider a distribution concentrated in just one point, namely, a Dirac delta
distribution).  Therefore, let us focus, for the moment, on regular
distributions with well-defined probability density $p(x)$ everywhere and $D_q
= 3$ for all $q$.%
\footnote{ In rigorous mathematical terms, the needed regularity condition is
{\em absolute continuity} with respect to the Lebesgue measure, namely, the
condition that every set with zero volume (null Lebesgue measure) contains no
mass.  It implies, by the Radon-Nikodym theorem, that the mass distribution is
given by the integral of a density that is unique (almost everywhere)
\cite{measure}.  In fact, absolute continuity allows some singularities, for
example, isolated power-law singularities. These singularities are compatible
with $D_1 = 3$, which is the only condition that we actually need in the
following.  Moreover, there are very mild singularities that are compatible
with $D_q = 3$ for all $q \in \mathbb{R}$.}
The probability in an element of volume $V$ is given, as $V \to 0$, by $p(x)
  V$, where $x$ belongs to that element of volume (this dependence of
  probability on volume derives from the local dimension being $\a=3$
  everywhere).  Therefore,
\begin{eqnarray*}
S(\{p_i\}) &\approx& -\sum_i p(x_i)\,V \log_2[p(x_i)\,V] \\
&=& -\sum_i p(x_i)\,V \log_2[p(x_i)] - \log_2 V,
\end{eqnarray*}
where the sum runs over a partition of the total volume in volume-$V$ elements
(a partition in cells, for example).  In the limit $V \to 0$, we can write the
entropy as the sum of a finite part and a divergent part, namely,
$$
S[p(x)]  \approx - \int p(x)\, d^3x\, \log_2[p(x)] - \log_2 V.
$$ 
Naturally, the divergent part just tells us that $D_1 = 3$, whereas the 
finite part is a non-trivial integral of the density.

The finite part of the total entropy is not defined in an absolute way: for
partitions in unequal volume elements, when the continuum limit is taken, the
logarithm in the integrand is replaced with $\log_2[\phi(x)p(x)]$, where
$\phi(x)$ is a positive function.  On the other hand, while the total entropy
is always positive, its finite part can be negative.  For these reasons, it is
necessary to introduce the {\em relative entropy}.  Conventionally, the
entropy of the density $p(x)$ relative to the density $q(x)$ is defined as%
\footnote{Here we incur a slight notational inconsistency, since we have been
  using $q$ for the parameter in the R\'enyi entropies or dimensions. Hence,
  we leave it to the reader to discern from the context whether $q$ means the
  probability distributions $q(x)$ or $q_i$ or the number $q$.}
$$
S(p | q) = \int p(x)\, d^3x\, \log_2\frac{p(x)}{q(x)}\,,
$$
where it is understood that $p(x)=0$ wherever $q(x)=0$.  The relative entropy
is always positive.  It is also called the Kullback or Kullback-Leibler
divergence, and it is studied in detail by Kullback \cite{Kullback} (note that
``divergence'' means discrimination measure in the statistical context).
Therefore, the absolute entropy of a coarse-grained distribution gives rise,
in the continuum limit, to an absolute part, the dimension, and a relative
part, the relative entropy.%
\footnote{It is useful (but optional) to also define the relative entropy of
  discrete distributions \cite{Renyi}.}  
Only the latter differentiates regular distributions.  Notice that the
entropy relative to the uniform distribution is simplest but is only
defined for distributions over a finite volume (in our case, the unit cube).%
\footnote{The relative entropy with respect to the uniform distribution has
  been considered as a measure of the evolution of inhomogeneity in cosmology
  by Hosoya, Buchert \& Morita \cite{Hosoya}.}

These results hold for singular multifractal distributions with $D_1 < 3$,
after the necessary adaptations. One singular distribution $\nu$ can be
relatively regular, that is to say, it can be regular with respect to another
singular distribution $\mu$.%
\footnote{Again, the appropriate mathematical definition of regularity
is absolute continuity, now with respect to the measure $\mu$ (every set with
null $\mu$-measure has null $\nu$-measure). By the Radon-Nikodym theorem,
there is a density $d\nu/d\mu$, unique except in a set of null $\mu$-measure.}
This essentially means that the singularities of $\nu$ form a subset of 
the singularities of $\mu$. 
The entropy of $\nu$ relative to $\mu$ is defined as
$$
S(\nu | \mu) = \int d\nu(x)\, \log_2\frac{d\nu(x)}{d\mu(x)} \geq 0, 
$$ where $d\nu(x)/d\mu(x)$ is the density of $\nu$ with respect to $\mu$ at
the point $x$.  This relative entropy differentiates one multifractal
distribution ($\nu$) from another ($\mu$), when the former is regular with
respect to the latter and, in particular, they have the same dimension $D_1$.
In fact, $S(\nu | \mu) = 0$ if and only if $\nu = \mu$.

The R\'enyi entropy $S_q$ (\ref{Sq}) also gives rise in the continuum limit to
a divergent part, and hence the dimension $D_q$, and to a finite part.  This
finite part motivates the definition of the relative R\'enyi entropy
\begin{equation*}
S_q(\nu | \mu) = \frac{1}{1-q}\,\log_2 \left[\int d\nu(x)
\left(\frac{d\nu(x)}{d\mu(x)}\right)^{q-1}\right], \quad 
q \neq 1\,.
\end{equation*}
However, this relative entropy is less useful than the 
standard (Kullback-Leibler) relative entropy. 

The relative entropy differentiates distributions but has two shortcomings.
First, $S(\nu | \mu)$ is only defined when $\nu$ is $\mu$-regular. Second, the
relative entropy does not have the necessary properties to qualify as a
distance between distributions: it fails to be symmetric or to fulfill the
triangle inequality.  However, it is possible to define a real distance
between any two distributions in terms of their entropies.  For discrete
distributions, Endres \& Schindelin \cite{IEEE} define
\begin{eqnarray*} 
D^2_{PQ} &=& 2 S(R) - S(P) - S(Q) \\
&=& \sum_{i=1}^M \left(p_i \log_2 \frac{2p_i}{p_i+q_i} 
 + q_i \log_2 \frac{2q_i}{p_i+q_i} \right),
\end{eqnarray*}
where $P=\{p_i\}$, $Q=\{q_i\}$ and $R=\{(p_i+q_i)/2\}$. Then, they prove that
$D_{PQ}$ is a distance.  Furthermore, Endres \& Schindelin \cite{IEEE}
note that it can be applied to continuous distributions. This follows from
the alternative expression
$$
D^2_{PQ} = S(P|R) + S(Q|R),
$$ 
that is to say, from $D^2_{PQ}$ being a sum of relative entropies, 
in addition to the fact
that any two continuous distributions are both regular
with respect to their mean.  Therefore, $D_{PQ}$ is well defined in the
continuum limit of $P$ and $Q$.

Thus, we can measure the distance between the coarse-grained distributions
$p_i = n_{{\rm g}\,i}/N$ and $q_i = n_{{\rm m}\,i}/N$, where $n_{{\rm
    g}\,i},n_{{\rm m}\,i} \gg 1$, and then we can take the continuum
limit. The distribution $R$ corresponds to the total particle distribution.
The squared distance between the coarse distributions is
\begin{eqnarray}
D^2_{PQ} = 
2+\frac{1}{N} \sum_{i=1}^M \left({n_{{\rm g}\,i}} \log_2 {n_{{\rm
g}\,i}} +{n_{{\rm m}\,i}} \log_2 {n_{{\rm m}\,i}} - 
(n_{{\rm m}\,i}+n_{{\rm
g}\,i}) \log_2(n_{{\rm m}\,i}+n_{{\rm g}\,i}) \right) 
\label{Dpq1}
\\
= \frac{1}{N} \sum_{i=1}^M \left({n_{{\rm g}\,i}} \log_2 {n_{{\rm
g}\,i}} +{n_{{\rm m}\,i}} \log_2 {n_{{\rm m}\,i}} + 
(n_{{\rm m}\,i}+n_{{\rm g}\,i}) 
[1-\log_2(n_{{\rm m}\,i}+n_{{\rm g}\,i})] \right).
\label{Dpq2}
\end{eqnarray}
Referring to the expression (\ref{mix}) of the cell entropy, we deduce that,
in the sum of terms (one per cell) given by Eq.~(\ref{Dpq2}), each term
represents the gap between the maximum cell entropy of mixing (one bit per
particle) and its actual value, just like in the sum of cell contributions in
the Bayes information (\ref{Bayes_info}).  Naturally, $D^2_{PQ}$ decreases
with mixing and vanishes for the most mixed distribution $P = Q =
R$. Conversely, it takes its maximum, $D^2_{PQ} = 2$, when $\{n_{{\rm
g}\,i}\}$ and $\{n_{{\rm m}\,i}\}$ are disjoint, namely, when they are not
mixed at all [as we deduce from Eq.~(\ref{Dpq1})].  Regarding the continuum
limits of $P$ and $Q$, Endres \& Schindelin's distance is maximal if they are
{\em mutually singular}, namely, if they concentrate in disjoint sets. The
continuum limits of disjoint $\{n_{{\rm g}\,i}\}$ and $\{n_{{\rm m}\,i}\}$
give rise to two mutually singular distributions but the definition
encompasses more general cases.%
\footnote{The definition of mutually singular distributions is given by, e.g.,
  Capinski \& Kopp \cite{measure}.  A particularly clear case of mutually
  singular distributions occurs when they have disjoint supports, but this is
  not necessary: for example, the uniform distributions in the Cantor set and
  in the unit interval, respectively, are mutually singular, although the
  Cantor set is contained in the unit interval.}

Let us notice that the above defined statistical distance is consistent with
our Bayesian analysis but cannot replace it.  Firstly, it relies on the
approximation $n_{{\rm g}\,i},n_{{\rm m}\,i} \gg 1$, that is to say, on
neglecting the discreteness effect due to particle counts.  In this
approximation, the entropy of mixing in the form given by Eq.~(\ref{mix}) is
just the asymptotic form of the cell entropies in Eq.~(\ref{Bayes_info}); but
note that the global contribution in Eq.~(\ref{Bayes_info}) diverges as $N \ra
\infty$.  Lastly, it is a general fact that a statistical distance cannot
provide a sharp criterion to decide if two discrete distributions are samples
from the same continuous distribution, and it is on the same footing as the
cross-correlation coefficient in that regard.

Endres \& Schindelin's distance can be connected with a standard statistical
measure of discrimination as follows.  Let us note that $D^2_{PQ}$ adopts a
simplified form when $P$ and $Q$ are close \cite{IEEE}, namely,
$$ D^2_{PQ} \approx \frac{1}{\ln 2}\sum_{i=1}^M \frac{(p_i - q_i)^2}{2(p_i
+ q_i)} = \frac{1}{2\ln 2}\,\chi^2_{PQ}\,,
$$
where the last expression refers to Pearson's chi-square test of
discrimination, which can be considered a particular case of the 
Endres-Schindelin distance.%
\footnote{The connection of Pearson's chi-square test with information theory
  can be obtained directly from the relative entropy \cite{Kullback}. However,
  $\chi_{PQ}$ is much closer to Endres \& Schindelin's distance: it is also a
  distance and, furthermore, $\chi^2_{PQ}/(2\ln 2) \leq D^2_{PQ} \leq
  \chi^2_{PQ}\,,$ for any $P$ and $Q$.}
In our case, 
$$ \chi^2 = \sum_{i=1}^M \frac{(n_{{\rm m}\,i} - n_{{\rm
g}\,i})^2}{n_{{\rm m}\,i}+n_{{\rm g}\,i}}\,.
$$ 
The chi-square test has the advantage of highlighting that the expected
fluctuations of $|n_{{\rm m}\,i} - n_{{\rm g}\,i}|$ in a common distribution
are of the order of $(n_{{\rm m}\,i} + n_{{\rm g}\,i})^{1/2}$.  At any rate,
the test is based on an approximation of $D^2_{PQ}$ and neither can it
provide a sharp criterion of discrimination.

\subsection{Bias as entropic distance}

In cosmology, the bulk of mass belongs to the dark matter, so the distribution
of gas (or galaxies) is assumed to be ``biased'' with respect to the total
matter distribution, dominated by the dark matter.  Since we normalize to one
both the dark matter and the gas total masses, both components play a
symmetrical r\^ole in our statistical analyses. Therefore, our measure of bias
must be just a measure of discrimination between two probability distributions
(a ``divergence'' or distance).  There are many such measures, but the notions
of relative entropy and Endres-Schindelin distance naturally arise in
connection with our Bayesian analysis.  Regarding the Endres-Schindelin
distance, mutually singular distributions are most distant, namely, at
distance $\sqrt{2}$.  This distance diminishes if the distributions
concentrate in a common set, but vanishes only when they coincide.  The
relative entropy is not a distance but it is useful as well, because it
diverges for mutually singular distributions and, therefore, it separates
distributions better.  In fact, the relative entropy can be symmetrized with
respect to the compared distributions, and then it diverges unless they are
mutually regular.%
\footnote{The symmetric relative entropy $S(P|Q) + S(Q|P)$ is called the
Jeffreys divergence $J(P,Q)$ \cite{Renyi,Kullback}. Despite being symmetrical,
it is not a proper distance, for it still fails to fulfill the triangle
inequality.  It is trivially finite for distributions that are mutually
regular, namely, absolutely continuous with respect to one another.  Kullback
\cite{Kullback} always works within an equivalence class of mutually regular
distributions.}

The simplest example of comparison of two distributions occurs when they are
both regular, in particular, when they have everywhere well-defined densities
$p(x)$ and $q(x)$. In spite of their individual regularity, they are mutually
singular if they do not overlap, that is to say, if each density is positive
only where the other density vanishes, then being at Endres-Schindelin
distance $\sqrt{2}$.  As they overlap more and, furthermore, the densities
approach one another, their Endres-Schindelin distance and their symmetric
relative entropy tend both to zero.  On the other hand, the symmetric relative
entropy is finite only if both distributions vanish in the same point set
(disregarding sets of zero volume, of course).

Regarding singular distributions, the first condition for two distributions to
be at small Endres-Schindelin distance is that they have the same R\'enyi
dimensions and, therefore, the same multifractal spectrum.  However, this
condition is far from being sufficient.  Indeed, the multifractal spectrum
only gives the ``size'' (the dimension) of every set of singularities with
common strength (local dimension), but tells us nothing about the precise
geometry (location or shape) of those sets.  Like in the case of regular
distributions, two distributions are at small Endres \& Schindelin's distance
if the strength and location of their mass concentrations, in particular,
their singularities, essentially coincide.  As regards the symmetric relative
entropy, the singularities must actually coincide for it to be finite.

It has been remarked above that a statistical distance (or divergence) cannot
provide a sharp distinguishability criterion.  In fact, the distinguishability
criterion provided by the Bayes factor only makes sense for finite point
distributions, namely, for deciding if two finite point distributions can be
samples from the same multinomial distribution.  In this regard, the Bayesian
comparison of the dark-matter and gas cell distributions in Sect.\
\ref{appl_Bayes} has clearly ruled out a common multinomial distribution on
nonlinear scales.  Nevertheless, the entropy of mixing per particle is very
close to the maximum of one bit; for example, it is $0.992$ bits for massive
halos in the master cell distributions.  Therefore, the two distributions are
indeed very mixed (very close).

Furthermore, the closeness of the gas and dark matter distributions suggests
that their individual singularities coincide and, therefore, the two
distributions are mutually regular.  In the coarse formalism that we use, the
local dimension of cell $i$ is
$$
\a_i = 3\frac{\log[n_i/(NV_0)]}{\log(V/V_0)}\,.
$$
Therefore, the difference between the strenghs of gas and dark matter
singularities is 
$$
\a_{{\rm g}\,i} - \a_{{\rm m}\,i}
= 3\frac{\log(n_{{\rm g}\,i}/n_{{\rm m}\,i})}{\log(V/V_0)}\,.
$$
We can see that this difference vanishes if $n_{{\rm g}\,i}/n_{{\rm m}\,i}$
stays bounded (above and below) while the cell volume $V$ shrinks.  Although
we have found that the ratio $n_{{\rm g}\,i}/n_{{\rm m}\,i}$ is not unity in
populated cells, its logarithm is small (in absolute value) with respect to
$-\log(V/V_0)$ at the lower end of the multifractal scaling range, thus making
$\a_{{\rm g}\,i}$ and $\a_{{\rm m}\,i}$ almost equal.  In general, if we
define a local bias factor as the local relative gas concentration, the
condition for common gas and dark-matter singularities is mild: the local bias
factor must be bounded away from zero and infinity.

\section{Discussion and Conclusions}
\label{discuss}

We have improved the method of coarse multifractal analysis based on counts in
cells by devising a procedure for extracting from a sample of a distribution
the maximal information about its multifractal properties. The procedure is
based on a clear understanding of the r\^ole of the upper and lower cutoffs to
scaling, which are, respectively, the homogeneity and discreteness scales.
The homogeneity scale is used in the definition of coarse multifractal
exponents [Eq.~(\ref{ctauq})], while the discreteness scale is crucial to
understand and quantify the effects of under-sampling.  We have employed our
procedure to analyse the gas and dark matter distributions in the Mare-Nostrum
universe at redshift $z=0$.

The only intrinsic scale present in an $N$-body simulation is actually the
discreteness scale $V= N^{-1}$ (besides the size of the simulation cube, which
we take as the reference scale).  The homogeneity scale is present as well but
it is dynamical and grows with time.  Between these two scales the matter
distribution can be considered continuous and representative of the nonlinear
dynamics.  The discreteness scale $V= N^{-1}$ defines what we call the master
cell distribution, which best resolves the overall mass distribution. The mass
function of objects at this scale (halos) adopts a power-law form with a
large-mass cutoff, similar to the Press-Schechter mass function. However, its
power-law exponent is $-2$, which would correspond to an initial power
spectrum with index $n=-3$ in the Press-Schechter theory, whereas the actual
value in the Mare-Nostrum universe is $n=1$. In conclusion, the Mare-Nostrum
mass function confirms the form of the mass function found in Ref.~\cite{I4}
and its independence of the initial power spectrum.

Of course, the Press-Schechter theory and the consequent mass function are not
applicable to equal-size objects.  However, Vergassola et al \cite{V-Frisch},
in their study of the adhesion model (described in Ref.~\cite{Shan-Zel}), also
define coarse-grained objects of equal size and, nevertheless, they find a
power-law mass function with exponent depending on the initial spectral index
and with an exponential large-mass cutoff, like in the Press-Schechter theory.
On the other hand, Vergassola et al \cite{V-Frisch} show that the adhesion
model gives rise to a multifractal cosmic-web structure (see also
Ref.~\cite{Bou-M-Parisi}).  In this regard, it is especially interesting to
compare our results with theirs, and to emphasize that the power-law exponent
$-2$ is unrelated to the initial power spectrum, unlike their power-law
exponent.  The dependence of their power-law exponent on the initial power
spectrum is surely due to the nature of the Zel'dovich approximation, in which
the dynamics is trivial before the formation of singularities.  In contrast,
the real gravitational dynamics is {\em chaotic}.  Therefore, the multifractal
attractor of the real dynamics is independent of the initial conditions and
must arise even when the initial conditions do not have a scale invariant
power spectrum.

The mass function power-law exponent $-2$ is, in fact, naturally associated
with the multifractal mass concentrate. Furthermore, we find that the precise
form of the exponential large-mass cutoff suggests that the power law is
actually an approximation of a lognormal mass function, as expected in a
multifractal \cite{I4} and found in the Mare-Nostrum universe on larger
scales.

Our first direct test of scale invariance consists in calculating the coarse
multifractal spectrum in a range of nonlinear scales, namely, from $l=2^{-12}$
up to $2^{-7}$.  For this, we use the improved definition of coarse
exponents~(\ref{ctauq}), which includes the scale of homogeneity (estimated
through the condition $\mu_2 =1.1$).
This improvement is necessary when the scale of homogeneity is considerable
smaller than the box size.  The resulting multifractal spectra (Fig.\
\ref{MFspec}) agree in their respective ranges (except near $\a_{{\rm
max}}$). Moreover, the spectra corresponding to the dark matter and to the gas
are almost identical.  However, the introduction of the scale of homogeneity
produces an anomalous extension of the multifractal spectrum: it gives rise to
{\em negative} fractal dimensions. They can be understood as representing
improbable matter fluctuations that can be ignored.

From the multifractal spectra, we deduce two important dimensions, namely, the
dimension of the mass support $D_0 = 3$ and the dimension of the mass
concentrate $D_1 \simeq 2.4$.  Both dimensions provide information on the type
of multifractal cosmic-web structure.  The former dimension shows that this
multifractal is non-lacunar while the latter shows that it is not very
concentrated.  The overall weak concentration indicated by $D_1 \simeq 2.4$
can be due to the dominance of surface singularities (``pancakes'') but can
also be due to the clustering of lower dimensional singularities, namely,
filaments or nodes.  Cosmic web singularities are difficult to define in
galaxy or $N$-body samples, but can be partially unveiled with appropriate
algorithms \cite{S4,vW-Sch}.  At any rate, one must notice that a non-lacunar
cosmic web structure has a very complex geometry \cite{I5}.  Of course, this
geometry is determined by the dynamics of gravitational collapse and, in
particular, by its type of anisotropy; but further discussion of this question
is beyond the scope of this work (the r\^ole of anisotropic collapse in the
formation of the cosmic web is discussed in Ref.~\cite{Rien}, for example).


Our study of the multifractal spectra on decreasing scales from $l=2^{-7}$ to
$2^{-12}$, including the discreteness scale $l= N^{-1/3}= 2^{-10}$, allows us
to discern the progressive influence of discreteness. The most obvious change
is, of course, the shrinking range of $\a$, namely, the reduction of $\a_{{\rm
max}}$ caused by lack of mass resolution: depleted small cells must be
empty. Furthermore, the mass distribution is under-sampled in cells with few
particles, altering the ends of the spectra near $\a_{{\rm max}}$.  We can
measure these deviations, for we can compare small scale spectra with 
the complete spectra at $l=2^{-7}$.
Actually, the spectra are almost complete at $l=2^{-8}$. For $l>2^{-7}$, 
there appear early signs of the transition to homogeneity.

It is interesting to connect our results about the influence of discreteness,
which only concern the statistical properties of the redshift $z=0$
distributions, with the studies by Kuhlman, Melott \& Shandarin \cite{KMS} and
Splinter et al \cite{KMSS} of the {\em dynamical} effects of
discreteness. Those authors conclude that these effects are the more important
the less converging the particle motion is. Thus, we have, on the one hand,
that expanding volume elements give rise to voids, with local dimension $\a >
3$, which are only well represented in the multifractal spectra corresponding
to scales considerably larger than $l = N^{-1/3}$. On the other hand,
collapsing volume elements give rise to mass concentrations with the smaller
dimension the larger is the number of independent axis along which they
collapse.  These mass concentrations can be well represented in the spectra
corresponding to $l < N^{-1/3}$.  For example, isotropic collapse gives rise
to the smallest dimension concentrations, which are the most robust against
the effects of undersampling; and, in fact, the low-$\a$ end of the
multifractal spectrum is essentially correct even for scales $l <
2^{-12}$. However, the strong singularities with low $\a$ do not represent the
full cosmic web structure.

Our second and most direct test of scale invariance is made in the standard
way, namely, by studying the dependence of the second order moment $M_2$ on
the scale $l$: we calculate $M_2(l)$ from $l=2^{-12}$ to $2^{-2}$, a broad
range that includes the discreteness and homogeneity scales. On the smaller
scales, we correct for the effect of discreteness by suppressing under-sampled
cells, according to the information provided by the already computed
spectra. We find two well-defined scaling ranges: the fractal range, spanning
from $l=2^{-12}$ to $2^{-6}$, and the homogeneous range, from $l=2^{-4}$
upwards. The transition to homogeneity takes place between $l=2^{-6}$ and
$l=2^{-4}$.  For definiteness, we choose as homogeneity scale $l_0=2^{-5}$,
which in physical units is 16 $h^{-1}$ Mpc.  The fractal correlation
dimensions are $D_2=1.26$, for the dark-matter, and $D_2=1.30$, for the gas,
in accord with conventional values of the galaxy correlation dimension
\cite{Jones-RMP,Sylos-Pietro}.

To find out if the equivalence of the gas and dark matter distributions goes
beyond their scaling properties, we have undertaken a detailed statistical
study of the relation between these distributions.  Since we employ the method
of counts in cells, we have specified two kinds of comparison: (i) the two
cell distributions, defined by their respective sets of occupation numbers
$\{n_i\}$, are compared as if they were two discrete probability distributions
with respective probabilities $\{p_i = n_i/N\}$; (ii) the two cell
distributions $\{n_{{\rm m}\,i}\}$ and $\{n_{{\rm g}\,i}\}$ are compared to
decide if it is likely that they are samples from the same multinomial
distribution (given by some coarse distribution $\{p_i\}$).  The first kind of
comparison leads us to measures discriminating between discrete probability
distributions (and between their continuum limits).  We have considered
firstly the cross-correlation coefficient and lastly entropic distances (or
``divergences''), actually motivated by our method of deciding if two cell
distributions are samples of the same multinomial distribution.  Since there
are many (pseudo)distances to discriminate between discrete probability
distributions, the comparison based on one of them has no absolute value.
However, all the measures that we employ to discriminate between the coarse
gas and dark matter distributions tell us that they are very close.

To decide if it is likely that the two cell distributions $\{n_{{\rm m}\,i}\}$
and $\{n_{{\rm g}\,i}\}$ are samples from the same multinomial distribution,
we develop a Bayesian method of analysis.  The two distributions are compared
by means of the Bayes information about the equality $p_{{\rm m}} = p_{{\rm
g}}$, namely, by means of the logarithm of the corresponding Bayes
factor~(\ref{Bf1}).  The Bayes information corresponding to a set of massive
cells can be expressed as a sum of negative cell terms, proportional to the
entropy of mixing per particle minus one, added to a positive global term.
The application of this formula to the master cell distributions, starting
from the most massive halos, demonstrates gas biasing.  In particular, the gas
is less concentrated in massive halos.  The bias is attenuated on larger
scales but only disappears at $l=2^{-4}$, namely, at the scale of full
homogeneity.  Naturally, it is to be expected that there is no bias at
homogeneity, for it essentially preserves the initial conditions.  However, we
do not have any argument that forbids that the bias vanishes at a smaller
scale, so the fact that it vanishes only at homogeneity could be coincidental.

Since the Bayesian analysis can be formulated in terms of the entropy of
mixing, we have studied in detail the entropic comparison of continuous
distributions.  We must assume that the R\'enyi entropies of the compared
distributions have well defined continuum limits, which amounts to assuming
that the distributions are multifractal (including regular distributions with
$D_q=3$).  Thus, the first element of comparison is the spectrum of R\'enyi
dimensions or, equivalently, the multifractal spectrum.  As regards their
multifractal spectra, the dark matter and gas distributions in the
Mare-Nostrum universe are indistinguishable.  However, the multifractal
spectrum gives the sizes of the sets of dark-matter or gas concentrations (or
depletions) with equal strength but is insensitive to the location of those
sets.  In fact, the R\'enyi dimensions only contain partial information about
a continuous distribution.  In particular, $D_1$ represents only one part of
its entropy.  Another part of the entropy is of relational nature and can be
expressed as a relative entropy or as a statistical entropic distance equal to
(the square root of) the {\em neg-entropy} of mixing, proportional to one
minus the entropy of mixing per particle.  The high entropy of mixing or small
entropic distance between the gas and dark-matter distributions is due to the
fact that their respective singularities actually coincide, namely, the
respective singularities at the same positions have equal local dimensions.

The appearance of common singularities in the gas and in the dark matter
surely has a physical origin, despite the differences between the dynamics of
each component.  It is natural to conjecture that the common multifractal
structure is due to the fact that the gas and the dark matter are both
dominated, on a long range of scales, by the gravitational interaction, which
produces common power-law singularities.  The differences in the dynamics are
the cause of gas biasing but do not interfere with the essential multifractal
features of the distributions (except on very small scales).  In fact, the
Mare-Nostrum universe is not based on a very realistic model of gas dynamics,
insofar as it does not consider thermal radiation or conduction.
Nevertheless, if the cosmic web singularity structure is due to gravity only,
the analysis of future simulations will corroborate that the gas biasing does
not alter that structure.  Then, we can speak of a kind of {\em universality}:
the cosmic dynamics has a unique type of cosmic web multifractal attractor,
independent of the initial conditions.  In particular, the multifractal
spectrum obtained here from the Mare-Nostrum universe or before from the GIF2
simulation \cite{I4} must be characteristic of the cosmic web.

\begin{acknowledgments}
I thank Gustavo Yepes for making the Mare-Nostrum data available to me. 
\end{acknowledgments}

\end{document}